\documentclass[aps,prd,twocolumn,preprintnumbers,superscriptaddress,showpacs]{revtex4}

\usepackage{amsmath,amssymb}
\usepackage[dvips]{graphicx}
\usepackage{bm}

\DeclareMathOperator{\tr}{tr}

\DeclareMathOperator{\eff}{eff}

\DeclareMathOperator{\ring}{ring}

\begin{document}

\preprint{\normalsize UT-Komaba/07-15} 

\title{
First-order restoration of SU($N_f$)$\times$SU($N_f$) chiral symmetry 
with large $N_f$ \\ and Electroweak phase transition
}


\author{Yoshio Kikukawa}
\email[]{kikukawa@hep1.c.u-tokyo.ac.jp}
\affiliation{Institute of  Physics, University of Tokyo\ 
Tokyo 153-8092, Japan}

\author{Masaya Kohda}
\email[]{mkohda@eken.phys.nagoya-u.ac.jp}
\affiliation{Department of Physics, Nagoya University\ 
Nagoya 464-8602, Japan}

\author{Junichiro Yasuda}
\email[]{yasuda@hep1.c.u-tokyo.ac.jp}
\affiliation{Institute of  Physics, University of Tokyo\ 
Tokyo 153-8092, Japan}


\date{\today}

\begin{abstract}
It has been argued by Pisarski and Wilczek that finite temperature restoration of the chiral symmetry 
SU($N_{f}$)$\times$SU($N_{f}$) is first-order for $N_{f} \geq 3$.  This type of chiral symmetry with 
a large $N_f$ may appear in the Higgs sector if one considers models such as walking technicolor theories. 
We examine  the first-order restoration of the chiral symmetry from the point of view of the electroweak phase transition. The strength of the transition is estimated in SU(2)$\times$U(1) 
gauged linear sigma model by means of the finite temperature effective potential at one-loop with 
the ring improvement.
%
Even if the mass of the neutral  
scalar boson corresponding to the Higgs boson
is larger than 114 GeV, 
the first-order transition can be strong enough for the electroweak baryogenesis, 
as long as the extra massive scalar bosons (required for the linear realization) 
are kept  heavier than the neutral scalar boson. 
Explicit symmetry breaking terms
reduce the strength of the first-order transition, but the transition can remain strongly first-order 
even when the masses of pseudo Nambu-Goldstone bosons become as large as  the current 
lower bound of direct search experiments.

\end{abstract}


\maketitle






\section{Introduction}
\label{sec:introduction}
The standard model (SM) fulfills all three Sakharov conditions \cite{Sakharov:1973} for generating a baryon asymmetry in the Universe \cite{Kuzmin:1985mm,Cohen:1990py,Cohen:1990it}. However, 
the model fails to explain the value of the asymmetry,  
$n_B/s = (8.7\pm 0.3)\times 10^{-11}$, measured through the Cosmic Microwave Background \cite{Spergel:2006hy},  
or the value required for the primordial nucleosynthesis \cite{Steigman:2005uz},  for two reasons.
The first reason is that CP violation from the Kobayashi-Maskawa mechanism \cite{Kobayashi:1973fv} 
is highly suppressed \cite{Shaposhnikov:1987tw,Farrar:1993sp,Farrar:1993hn,Gavela:1994dt,Gavela:1993ts,Huet:1994jb}. The second reason is that the electroweak phase transition (EWPT) is not strongly first-order. The experimental lower bound on the Higgs mass,  $m_H > 114$ GeV \cite{:2003ih}, implies that there is no EWPT in the SM \cite{Kajantie:1996mn,Rummukainen:1998as,Csikor:1998eu,Aoki:1999fi}. Consequently, spharelon-induced (B+L)-violating interactions are not sufficiently suppressed in the broken phase and wash out the baryon asymmetry. 
If the physics at the electroweak scale could explain the baryon asymmetry in the Universe,  the better understanding of  the structure of the Higgs sector and the source of CP violation would be required.

Various extensions of the Higgs sector have been investigated from the above point of view:  these include 
two Higgs doublet model \cite{Bochkarev:1990fx,Bochkarev:1990gb,Cohen:1991iu,Nelson:1991ab,Turok:1990zg,Turok:1991uc,Funakubo:1993jg,Davies:1994id,Cline:1995dg,Cline:1996mg,Fromme:2006cm, Kanemura:2004ch},  
minimal SUSY standard model (MSSM) \cite{Carena:1996wj,Cline:1996cr,Laine:1996ms,Losada:1996ju,Cline:1998hy,Laine:1998vn,Laine:1998qk,Cline:1997vk,Huet:1995sh,Carena:1997gx,Joyce:1999fw,Joyce:2000ed,Kainulainen:2001cn,Kainulainen:2002th,Funakubo:1998jz,Funakubo:1998fk,Funakubo:1999ws,Funakubo:2002yb}, 
MSSM with an extra singlet scalar \cite{Pietroni:1992in,Davies:1996qn,Huber:1998ck,Huber:2000mg,Kang:2004pp,Funakubo:2004ka,Funakubo:2005pu,Funakubo:2005bu,Huber:2006wf},  
the SM with a low cut-off \cite{Grojean:2004xa,Ham:2004zs,Bodeker:2004ws}, and so on. 
The orginal Higgs sector of the SM, if the electroweak interactions is turned off,  
is nothing but the O(4) linear sigma model and
its finite temperature phase transition is second order,  which is governed by the 
Wilson-Fisher IR-stable fixed point.
It is the effect of the gauge interaction which makes the fixed point IR-unstable and causes a 
first-order phase transition \cite{Coleman:1973jx}. 
Once the Higgs sector is extended, the number of the scalar fields is increased and there appear 
additional quartic couplings among them. The fixed points of the multiple quartic coupling constants 
may be IR-unstable and 
one can expect a first-order phase transition even in the pure scalar 
sector \cite{Bak-Krinsky-Mukamel:1976,Iacobson:1981jm}. 
A related approach is to consider the single Higgs doublet model (the O(4) model) with the dimension-six or higher operators \cite{Grojean:2004xa,Ham:2004zs,Bodeker:2004ws}. The quartic coupling is then assumed to be negative so that the model is out of the domain of the Wilson-Fisher  fixed point.  
Such higher dimensional operators may be induced by the effect of heavy particles coupled to the Higgs doublet, or more generally, by the effect of a certain dynamical system behind the Higgs sector. 

In this paper, we consider the finite temperature restoration of SU($N_f$)$\times$SU($N_f$) chiral symmetry with a large $N_f$ from the point of view of the electroweak phase transition.  
It has been argued by Pisarski and Wilczek that {\em if $N_f \ge 3$, the restoration of the chiral symmetry at finite temperature should be first-order}, through the renormalization group analysis of linear sigma model as a low energy effective theory of QCD \cite{Pisarski:1983ms} \cite{Rudnick:1978, Iacobson:1981jm, Amit:1978dk, Ginsparg:1980ef,Yamagishi:1981qq, Chivukula:1992pm,
Appelquist:1995en,Khlebnikov:1995qb}. 
The chiral symmetry with a large $N_f$ may appear in the Higgs sector as a symmetry of a certain underlying dynamical system such as walking technicolor theories \cite{Weinberg:1979bn,Susskind:1978ms,Eichten:1979ah,Holdom:1981rm,Holdom:1984sk,Yamawaki:1985zg,Appelquist:1986an,Appelquist:1986tr,
Appelquist:1987fc, Sannino:2004qp, Dietrich:2006cm}.
We will examine whether  the first-order phase transition associated with the chiral symmetry 
restoration can be strong enough for the electroweak baryogenesis to operate. 
A similar consideration has been performed by Appelquist et al. in \cite{Appelquist:1995en}. 

Following Pisarski and Wilczek, we consider the renormalizable SU($N_f$)$\times$SU($N_f$) linear sigma model with a large $N_f$ as a low energy effective theory of a certain underlying dynamical system such as walking techinicolor theories. In adopting the model for the Higgs sector,  we couple the SU(2)$\times$U(1) electroweak interactions
and introduce the explicit symmetry breaking terms which give rise to the masses of the extra 
Nambu-Goldstone (NG) bosons.
We will examine the chiral symmetry restoration in the model 
by the $\epsilon$-expansion \cite{Pisarski:1983ms, Rudnick:1978, Iacobson:1981jm, Amit:1978dk, Ginsparg:1980ef,Yamagishi:1981qq, Chivukula:1992pm} and 
by the finite temperature effective potential obtained  at one-loop in the ring-improved perturbation theory \cite{Dolan:1973qd,Weinberg:1974hy,Fendley:1987ef,Espinosa:1992gq,Parwani:1991gq,Anderson:1991zb,Carrington:1991hz,Dine:1992wr,Arnold:1992rz,Fodor:1994bs}. The sphaleron energy will be also re-examined within this model in order to clarify the criterion for the strength of the first-order  transition \cite{Manton:1983nd,Klinkhamer:1984di}. 
In \cite{Appelquist:1995en},  the finite-temperature effective potential 
has been examined in a large $N_f$ expansion, but only for  the case  of  
the pure linear sigma model 
without the gauge interactions nor the explicit symmetry breaking terms.  

Since the critical behavior of first-order phase transition at finite temperature  that we concern, is not universal in general, the result of our analysis would depend on our choice of low energy effective theory, where a certain truncation of fields and operators has to be done.
Moreover,  for the validity of the perturbation theory  at finite temperature, 
the gauge- and scalar- coupling constants must satisfy certain constraints. 
It turns out that one cannot push the masses of the scalar fields to so large values as expected in walking technicolor theories: 
the masses of the scalar fields are up to 300 GeV and  in particular 
the mass of the neutral scalar field which corresponds to the Higgs field is up to 150 GeV.  
Our analysis, therefore, must be qualitative,
just showing a possibility to realize strongly first-order electroweak phase transition required for the electroweak baryogenesis.  We leave more thorough quantitative study of this possibility for other non-perturbative methods like
the Monte Calro simulation in lattice gauge 
theory \cite{Iwasaki:2003de,AliKhan:2000iz,Damgaard:1997ut,Maezawa:2007fd,Ukita:2006pc}. 

We do not address,  in this paper,  the question about a  possible new source of CP violation which is required for the electroweak baryogenesis.  
The sector which is responsible for the flavor physics, in particular, the generation of the masses of quarks and leptons,  may well affect the dynamics of the chiral symmetry breaking/restoration. 
This effect of the flavor sector may be partly incorporated into the linear sigma model through the Yukawa couplings to quarks and leptons and the explict symmetry breaking terms which give rise 
to the masses and the interactions of the pseudo Nambu-Goldstone (NG) bosons.
However, the experimental constraints on the  flavor sector make it 
quite non-trivial to build a model of dymanical electroweak symmetry breaking which takes account of the flavor physics. (See \cite{Appelquist:2003hn} for recent progress in constructing a complete theory of fermion masses in the context of extended technicolor theories.)
In view of this situation, we 
simply omit the Yukawa couplings to quarks and leptons (and the CP violating phases) 
and consider  only the effect of the explicit symmetry breaking terms in the scalar fields. We choose the range and the pattern of the mass spectrum of the pseudo NG bosons so that the masses are within the allowed 
region of the direct search experiments \cite{Yao:2006px} 
and the contribution to the S parameter is not positive \cite{Dugan:1991ck,Peskin:2001rw}. 

At zero temperature, the phase transition associated with the breaking/restoration of 
SU($N_f$) $\times$ SU($N_f$) chiral symmetry  with a large $N_f$ has been examined by several authors \cite{Banks:1981nn,Appelquist:1996dq,Miransky:1996pd,Chivukula:1996kg,Appelquist:1998rb,Harada:2003dc,Kurachi:2006mu,Kurachi:2006ej}. In SU($N$) vector-like gauge theories with $N_f$ massless flavors, there appears a non-trivial IR fixed point (the Banks-Zaks fixed point) \cite{Banks:1981nn}  in the weak coupling perturbation theory for $N_f$ large, but  slightly less than $11N/2$.  As $N_f$ is lowered, 
the value of the fixed-point coupling becomes larger.  If the Banks-Zaks fixed point persists to larger coupling,  it may exceed the critical value for chiral symmetry breaking.  It has been argued that this "phase transition in the number of flavors $N_f$" shows the peculiar nature: the transition is not second order even though the order parameter changes continuously. The critical behavior is not universal,  where, from the side of the 
broken phase,  the entire spectrum collapses to zero mass and contributes to 
any correlation functions. 
Then, one cannot reduce the theory to an effective low-energy scalar theory.  
However, in a real life, $N_f$ is supposed to be a fixed integer between the "critical value" 
$N_f^c = N (100N^2-66)/(25N^2-15)$ and $11N/2$.  Although there would be some very light 
particles besides the NG bosons,  one may assume a description by a certain low energy effective theory which may include these light particles.
The linear sigma model we consider in this paper is hopefully one of such low energy effective 
theories \footnote{
In this respect, it may be useful to compare with the result  
using the low energy effective theory with light vector mesons such as the non-linear sigma model 
with hidden local symmetry \cite{Bando:1984ej,Bando:1987br,Harada:1993jk,Harada:1993qi,Harada:2003jx,Harada:2000kb,Harada:2001it}, which includes vector mesons instead of 
scalar mesons.}.

This paper is organized as follows. 
In section~\ref{sec:gauged-linear-sigma-model}, we describe how to introduce 
the SU(2)$\times$U(1) gauge interactions and the explicit symmetry 
breaking terms to the SU($N_f$)$\times$SU($N_f$) linear sigma model.  
In section~\ref{sec:epsilon-expansion}, we examine the (non-gauged) model by 
the $\epsilon$-expansion.  We will see that the model shows a first-order transition in the parameter region close to the "stability-boundary". 
Then we proceed to the analysis of the SU(2)$\times$U(1) gauged linear sigma model. 
In section~\ref{sec:sphaleron},  we examine the sphaleron energy. 
In section~\ref{sec:analysis-effective-potentail}, 
we obtain the finite temperature effective potential at one-loop with the ring-improvement and 
examine it both analytically (in high-temperature expansion) and numerically.  
Section~\ref{sec:summary-discussion} is devoted to a summary and discussions. 

\section{SU($N_f$)$\times$SU($N_f$) linear sigma model 
and the electroweak interactions\label{sec:gauged-linear-sigma-model}}

\subsection{SU($N_f$)$\times$SU($N_f$) linear sigma model}
\label{subsec:non-gauged-linear-sigma-model}

We consider the renormalizable linear sigma model which is defined by the following lagrangian:
\begin{eqnarray}
\mathcal{L}_\Phi &= &
\tr (\partial _\mu \Phi ^\dagger \partial ^\mu \Phi ) - m_{\Phi}^{2} \tr\Phi ^\dagger \Phi \nonumber\\
&& 
 - \frac{\lambda _1}{2} (\tr\Phi ^\dagger \Phi)^2 - \frac{\lambda _2}{2}\tr (\Phi ^\dagger \Phi )^2 . 
 \label{eq:Llsm}
\end{eqnarray}
The field $\Phi(x)$ is a $N_f \times N_f$ complex matrix which transforms under the chiral symmetry as 
\begin{equation}
\Phi \rightarrow  {\rm e}^{ i \alpha } \, g_L \Phi g_{R}^{-1} ,  \quad g_L, g_R   \in \text{SU($N_f$)} ; \ 
{\rm e}^{ i \alpha } \in U(1)_A . 
\end{equation}
For stability,  the quartic couplings should statisfy  the following conditions at tree level:
$\lambda_2>0,\lambda_1+\lambda_2/N_f>0$.  

We assume that the chiral symmetry breaks down to the diagonal subgroup SU($N_f$)$_V$ by 
the vacuum expectation value (VEV) of $\Phi(x)$:
\begin{equation}
\langle \Phi \rangle = \frac{\phi_0}{\sqrt{2N_f}} \openone, 
\end{equation}
where $\openone$ is the $N_f \times N_f$ unit matrix. 
At tree level, the VEV is determined by the effective potential: 
\begin{equation}
\label{eq:eff-potential-tree}
V_0(\phi)=\frac{1}{2}m_\Phi^2\phi^2+\frac{1}{8}\left( \lambda_1+\frac{\lambda_2}{N_f} \right)\phi^4 .
\end{equation}
For $m_{\Phi }^2 < 0$, it is given by 
\begin{equation}
\phi_0=\sqrt{\frac{-2m_\Phi^2}{\lambda_1+\lambda_2/N_f}}. 
\end{equation}

Around the VEV, we may parametrize the fluctuation of $\Phi(x)$ as follows:
\begin{equation}
\Phi(x) = \frac{ \phi + h + i \eta }{\sqrt{2N_f}}\, \openone
+ \sum_{\alpha=1}^{N_f^2-1} ( \xi^\alpha + i \pi^\alpha ) T^\alpha , 
\end{equation}
where $T^\alpha$ $(\alpha=1,\cdots, N_f^2-1)$ are the generator of SU($N_f$) with the normalization 
${\rm tr} (T^\alpha T^\beta)=\delta^{\alpha \beta}/2$. 
The fields $h, \eta, \xi^\alpha, \pi^\alpha$ acquire masses at the tree level as summerized in 
TABLE~\ref{tab:lsmm}, where, 
for notational simplicity, we use the following abbreviations:
\begin{eqnarray}
a_h&=&\frac{3}{2}(\lambda_1+\lambda_2/N_f), \\
a_\xi &=&\frac{1}{2}(\lambda_1+3\lambda_2/N_f),  \\
a_\eta&=&a_\pi=\frac{1}{2}(\lambda_1+\lambda_2/N_f), 
\end{eqnarray}
and
\begin{eqnarray}
b_h &=& a_h - a_\pi = (\lambda_1+\lambda_2/N_f), \\
b_\xi &=& a_\xi - a_\pi=(\lambda_2/N_f). 
\end{eqnarray}
$h$, the singlet of SU($N_f$)$_V$,  corresponds to the Higgs field. The adjoint $\pi^{\alpha}$ are 
the NG bosons of the breaking of SU($N_f$)$\times$SU($N_f$), while the singlet $\eta$ is 
the NG boson of the breaking of U(1)$_A$. The adjoint $\xi^{\alpha}$ are the extra massive degrees 
of freedom of the linear sigma model, which are required for the linear realization of the 
chiral symmetry.  Three of the NG bosons $\pi^{\alpha}$ are eaten by the SU(2)$\times$U(1) gauge bosons
when the electroweak interactions are introduced.

\begin{table}[t]
\begin{center}
  \begin{tabular}{cccc} \hline
    particle & $m_i^2(\phi)$ & $m_i^2(\phi_0)$ & $n_i$ \\ \hline
    h & $m_\Phi^2+a_h \phi^2$ & $b_h \phi_0^2$ & 1 \\
    $\xi$ & $m_\Phi^2+a_\xi \phi^2$ & $b_\xi \phi_0^2$ & $N_f^2-1$ \\
    $\eta$ & $m_\Phi^2+a_\eta \phi^2$ & 0 & 1 \\ 
    $\pi$ & $m_\Phi^2+a_\pi \phi^2$ & 0 & $N_f^2-1$ \\ \hline
\end{tabular}
\caption{The effective masses and the degrees of freedom of the fields in 
SU($N_f$)$\times$SU($N_f$) linear sigma model.}
\label{tab:lsmm}
\end{center}
\end{table}

For $N_f \le 4$, we may also add the relevant term, 
\begin{equation}
\label{eq:u1axial-breaking}
\mathcal{L}_{\Phi}^{\prime} = c^{\prime} (\det \Phi + \det \Phi ^\dagger ) , 
\end{equation}
which breaks the ${\rm U}(1)_{A}$ symmetry explicitly, but preserves CP symmetry. This term lifts 
the mass of $\eta$ to a nonzero value. 
In the case of $N_f=2$, the term is quadratic and the mass spectrum changes as in TABLE~\ref{tab:lsmm+cprime}. 
We note that  by sending $c^\prime$ to infinity one can decouple the field components
$\xi$ and $\eta$. In this case,  the original field $\Phi$ reduces to an O(4) variable, 
\begin{equation}
\Phi(x) =\frac{1}{2}\left[ (\phi + h)\, \openone
+ i \sum_{\alpha=1}^{3}  \pi^\alpha  \sigma^\alpha \right] ,  
\end{equation}
and  also 
the quartic couplings reduce to the single one in the combination $\lambda_1+\lambda_2/2$
due to the identity $\text{tr}(\Phi \Phi^\dagger)^2 = (\text{tr}\Phi \Phi^\dagger)^2/2$. 
Thus the model reduces to the O(4) linear sigma model.  For $N_f=3,4$, the term  contributes to
the effective potential as follows:
\begin{equation}
\label{eq:eff-potential-tree-det}
V_0(\phi)=\frac{1}{2}m_\Phi^2\phi^2+\frac{1}{8}\left( \lambda_1+\frac{\lambda_2}{N_f} \right)\phi^4 
- 2c^\prime \frac{\phi^{N_f}}{( \sqrt{2N_f})^{N_f}}. 
\end{equation}
The addition of the cubic term in the case of $N_f=3$ may enhance the first-order chiral phase transition. 
\begin{table}[t]
\begin{center}
  \begin{tabular}{cccc} \hline
    particle & $m_i^2(\phi)$ & $m_i^2(\phi_0)$ & $n_i$ \\ \hline
    h & $m_\Phi^2-c^\prime+a_h \phi^2$ & $b_h \phi_0^2$ & 1 \\
    $\xi$ & $m_\Phi^2+c^\prime +a_\xi \phi^2$ & $b_\xi \phi_0^2+2 c^\prime$ & $N_f^2-1$ \\
    $\eta$ & $m_\Phi^2+c^\prime+a_\eta \phi^2$ & 2 $c^\prime$ & 1 \\ 
    $\pi$ & $m_\Phi^2-c^\prime+a_\pi \phi^2$ & 0 & $N_f^2-1$ \\ \hline
\end{tabular}
\caption{The effective masses and the numbers of degrees of freedom in 
SU($N_f$)$\times$SU($N_f$) linear sigma model with the ${\rm U}(1)_{A}$ breaking term ($N_f$=2).}
\label{tab:lsmm+cprime}
\end{center}
\end{table}

\subsection{SU(2)$\times$U(1) gauged linear sigma model}

Next we consider the coupling of the electroweak interactions to the linear sigma model. 
SU(2)$\times$U(1) gauge interactions may be introduced through the minimal 
coupling to $\Phi(x)$ as 
\begin{eqnarray}
\label{eq:cov1d}
D_\mu \Phi &=&\partial_\mu\Phi -igA_\mu^a R(T^a_L) \Phi  \nonumber\\
&& \qquad - \, ig^\prime B_\mu \left[ R(Y_L)  \Phi - \Phi R(Y_R) \right], 
\end{eqnarray}
where $R(T^a_L)$ ($a=1,2,3$) and $R(Y_L), R(Y_R)$ are the generators of SU(2) and U(1), respectively,  represented by 2$\times$2 block-diagonal traceless hermitian matrices (assuming $N_f$ is even).  Here we consider two types of  representations inspired by the one-family model and the partially-gauged 
model \cite{Dietrich:2006cm} of technicolor theories:

\begin{table}[t]
\begin{center}
  \begin{tabular}{cccc} \hline
    particle & $m_i^2(\phi)$ & $m_i^2(\phi_0) $ & $n_i$ \\ \hline
    W & $\frac{g^2}{4}\phi^2$ & $\frac{g^2}{4}v^2$ & 6 \\
    Z & $\frac{g^2+g^{\prime 2}}{4}\phi^2$ & $\frac{g^2+g^{\prime 2}}{4}v^2$ & 3 \\ \hline
  \end{tabular}
\caption{The mass and the number of degrees of freedom of the gauge bosons in 
the one-family type of the SU(2)$\times$U(1) gauged linear sigma model.}
\label{tab:1fwm}
\end{center}
\end{table}

\begin{table}[h]
\begin{center}
  \begin{tabular}{cccc} \hline
    particle & $m_i^2(\phi)$ & $m_i^2(\phi_0) $ & $n_i$ \\ \hline
    W & $\frac{g^2}{2N_f}\phi^2$ & $\frac{g^2}{4}v^2$ & 6 \\
    Z & $\frac{g^2+g^{\prime 2}}{2N_f}\phi^2$ & $\frac{g^2+g^{\prime 2}}{4}v^2$ & 3 \\ \hline
\end{tabular}
\caption{The mass and the number of degrees of freedom of the gauge bosons in 
the partially-gauged type of the SU(2)$\times$U(1) gauged linear sigma model.}
\label{tab:pgwm}
\end{center}
\end{table}

\noindent (a) one-family type
\begin{eqnarray}
R(T^a_L) &=& \frac{\sigma^a}{2}  \, \otimes \, 
\text{diag}(1, \cdots, 1) ,  \\
R(Y_L) &=&  \, 0 , \\
R(Y_R) &=&  \frac{\sigma^3}{2}  \, \otimes \, \text{diag}(1, \cdots, 1) . 
\end{eqnarray}

\noindent (b) partially-gauged type
\begin{eqnarray}
R(T^a_L) &=& \frac{\sigma^a}{2}  \, \otimes \, \text{diag}(1, 0, \cdots) ,  \\
R(Y_L) &=&  \, 0 , \\
R(Y_R) &=&  \frac{\sigma^3}{2}  \, \otimes \, \text{diag}(1, 0, \cdots) . 
\end{eqnarray}

\noindent For these two types, 
the masses of the gauge bosons, $W^\pm$ and $Z$, are obtained as summerized in 
TABLE~\ref{tab:1fwm} and TABLE~\ref{tab:pgwm}, respectively. 
We note that in these two types,  there are different relations of  $\phi$ to the weak scale $v=246 \text{GeV}$.  Namely, we have 

\noindent (a) one-family type
  \begin{equation}
\phi_0 = v \qquad \text{[one-family type]}. 
\end{equation}  
\noindent (b) partially-gauged type
\begin{equation}
\label{eq:phi-v-rel-partiall}
\phi_0=\sqrt{\frac{N_f}{2}}\, v \quad \text{[partially-gauged type]}. 
\end{equation}

For the one-family type with $N_f = 8$, 
one may consider to couple the color gauge interaction. In this case, the 
minimal coupling Eq.~(\ref{eq:cov1d}) should be modified as follows: 
\begin{eqnarray}
\label{eq:cov1d+c}
D_\mu \Phi &=&\partial_\mu\Phi -igA_\mu^a R(T^a_L) \Phi  \nonumber\\
&& \qquad - \, ig^\prime B_\mu \left[ R(Y_L)  \Phi - \Phi R(Y_R) \right]  \nonumber\\
&& \qquad - \, ig_3 G_\mu^A  \left[ R(S^A) ,   \Phi    \right] , 
\end{eqnarray}
where
\begin{eqnarray}
R(T^a_L) &=& \frac{\sigma^a}{2}  \, \otimes \, 
\text{diag}(1, 1, 1, 1) ,  \\
R(Y_L) &=&  \, \openone_2 \, \otimes \, 
\text{diag}({\scriptstyle \frac{1}{6}},{\scriptstyle \frac{1}{6}} ,{\scriptstyle \frac{1}{6}}, {\scriptstyle -\frac{1}{2}}), \\
R(Y_R) &=&  \frac{\sigma^3}{2}  \, \otimes \, 
\text{diag}(1, 1, 1, 1) \ \oplus 
\nonumber\\
&&  \, \openone_2 \, \otimes \, 
\text{diag}({\scriptstyle \frac{1}{6}}, {\scriptstyle \frac{1}{6}},{\scriptstyle \frac{1}{6}}, {\scriptstyle -\frac{1}{2}}). 
\end{eqnarray}
and $R(S^A)$($A =1,\cdots,8$) are the generators of SU(3)$_c$ represented 
by the Gell-Mann matrices $\lambda^A$ as
\begin{equation}
R(S^A) = 
\openone_2 \, \otimes \, \left( \begin{array}{cc} \frac{\lambda^A}{2} & \vdots \\ \cdots & 0 \end{array}\right) . 
\end{equation}
Accordingly, the field $\Phi$ may be decomposed into the irreducible representations of SU(3). In terms of 
$2\times 2$ complex matrix valued fields, $\phi_S$, $\phi_{S^\prime}$, $\phi_{O}^A$($A =1,\cdots,8$),  $\phi_T^i$, $\phi_{\bar T}^i$($i=1,2,3$),  we have the following decomposition, 
\begin{equation}
\Phi=\left( \begin{array}{cc}  
\frac{1}{2} \phi_S \openone_3 + \frac{1}{2\sqrt{3}}\phi_{S^\prime}\openone_3+ \phi_{O}^A \frac{\lambda^A}{\sqrt{2}} 
& \phi_T\\
{}^t\phi_{\bar T} & \frac{1}{2} \phi_S - \frac{\sqrt{3}}{2} \phi_{S^\prime} \end{array} \right) . 
\end{equation}
where $\openone_3=\text{diag}(1,1,1)$. 

\subsection{Explicit symmetry breaking terms}
\label{subsec:symmetry-breaking}

In order to give rise to the masses to the extra $N_f^2-4$ NG bosons, we introduce the terms which break the chiral symmetry explicitly. 
For simplicity, we consider the minimal breaking for the one-family type, which preserves
SU(2)$_L$$\times$SU(2)$_R$$\times$SU($N_f/2$)$_V$ subgroup of the original chiral symmetry so that it is consistent with the gauge symmetries SU(2)$\times$U(1)($\times$ SU(3)). 
For this purpose, we may parametrize the fluctuation of $\Phi(x)$ by 2$\times$2 complex matrix 
valued fields as follows (assuming $N_f$ is even):
\begin{equation}
\Phi(x) = \phi_S(x) \otimes \sqrt{\frac{2}{N_f}}\, \openone_{N_f/2} 
+ \sum_{\alpha=1}^{(N_f/2)^2-1} \phi_P^\alpha(x)  \otimes S^\alpha , 
\end{equation}
and 
\begin{eqnarray}
\phi_S(x) &=&( \phi + h(x) + i \eta(x) )  \frac{\openone_2}{2}  + ( \xi_S^a(x)+ i \pi_S^a(x) ) \frac{\sigma^a}{2}, \nonumber\\
\phi_P^\alpha(x) &=& \xi_P^\alpha(x) + i \pi_P^\alpha(x)  \quad ( {\xi_P^\alpha}^\dagger = \xi_P^\alpha, {\pi_P^\alpha}^\dagger =\pi_P^\alpha), 
\end{eqnarray}
where $S^\alpha$ $(\alpha=1,\cdots, (N_f/2)^2-1)$ are the generator of SU($N_f/2$) with the normalization ${\rm tr} (S^\alpha S^\beta)=\delta^{\alpha \beta}$. 
$\phi_S(x)$ in the singlet of SU($N_f/2$)$_V$ and $\phi_P^\alpha(x)$ in the adjoint representation of SU($N_f/2$)$_V$ transform
under the SU(2)$_L$$\times$SU(2)$_R$ transformation as
\begin{eqnarray}
\phi_S \rightarrow   \, U_L \phi_S U_{R}^{-1} ,\ \phi_P^\alpha \rightarrow   \, U_L \phi_P^{\alpha} U_{R}^{-1}, 
\end{eqnarray}   
where $U_L, U_R \in$ SU(2). 
Then we consider the term 
\begin{equation}
\label{eq:symmetry-breaking}
\Delta{\cal L}_{sb} = \Delta {\cal L}_2+\Delta {\cal L}_4 , 
\end{equation}
where 
\begin{eqnarray}
\Delta {\cal L}_2&=&-\Delta m^2  {\rm tr} \{ \phi_S^\dagger \phi_S^{} \} +c \{ \det \phi_S^{} + c.c. \}  \nonumber\\
&& + \sum_{\alpha=1}^{(N_f/2)^2-1} c_P \{ \det \phi_P^\alpha + c.c. \}
\end{eqnarray}
and $\Delta {\cal L}_4$ is the quartic coupling term which consists of  ${\rm tr} \{ \phi_S^\dagger \phi_S^{} \}$, 
$\det \phi_S^{} $,  $\det \phi_P^\alpha$, ${\rm tr} \{ \Phi^\dagger \Phi \}$, and so on. 

For further simplicity, we restrict ourselves to the parameter region where  the last term of 
$\Delta {\cal L}_2$ and 
$\Delta {\cal L}_4$ vanish identically, including possible one-loop corrections,  and assume that the vacuum preserves SU($N_f/2$)$_V$.  
In this case, 
the coefficient of the quadratic term of the tree level effective potential Eq.~(\ref{eq:eff-potential-tree}) shifts:
$m_\Phi^2 \rightarrow m_\Phi^2+\Delta m^2-c$.  
The VEV is then given at tree level by 
\begin{equation}
\phi_0=\sqrt{\frac{-2(m_\Phi^2+ \Delta m^2 -c)}{\lambda_1+\lambda_2/N_f}}. 
\end{equation}
The masses of the fields $h, \eta, \xi_S, \pi_S, \xi_P, \pi_P$ at tree level 
are summerized in TABLE~\ref{tab:lsm+sb}.  To escape the vacuum instability, 
we require $c > 0$ and $c - \Delta m^2 > 0$ at tree level.  We note that
with this mass spectrum of the pseudo NG bosons (at tree level), the total constribution of the pseudo 
NG bosons to the S parameter is negative \cite{Dugan:1991ck,Peskin:2001rw}. 

\begin{table}[t]
\begin{center}
  \begin{tabular}{cccc} \hline
    particle & $m_i^2(\phi)$ & $m_i^2(\phi_0)$ & $n_i$ \\ \hline
    h & $m_\Phi^2+\Delta m^2-c+a_h \phi^2$ & $b_h {\phi_0}^2$ & 1 \\
    $\xi_S$ & $m_\Phi^2+\Delta m^2+c+a_\xi \phi^2$ & $b_\xi {\phi_0}^2 + 2 c $ & $3$ \\
    $\eta$ & $m_\Phi^2+\Delta m^2+c+a_\eta \phi^2$ & $2 c$  & 1 \\
    $\pi_S$ & $m_\Phi^2+\Delta m^2-c+a_\pi \phi^2$ & 0 & $ 3 $ \\ 
    $\xi_P$ & $m_\Phi^2+a_\xi \phi^2 $ & $b_\xi {\phi_0}^2+c - \Delta m^2 $ & $N_f^2-4$ \\
    $\pi_P$ & $m_\Phi^2+a_\pi \phi^2$ & $c - \Delta m^2$  & $N_f^2-4$ \\ \hline
\end{tabular}
\caption{
The effective masses and the numbers of degrees of freedom in 
SU($N_f$)$\times$SU($N_f$) linear sigma model with the symmetry breaking terms.}
\label{tab:lsm+sb}
\end{center}
\end{table}

\section{First-order restoration of  SU($N_f$)$\times$SU($N_f$) chiral symmetry with a large $N_f$ in $d=4-\epsilon$\label{sec:epsilon-expansion}} 

In this section, 
we examine the critical behavior of the chiral symmetry restoration at finite temperature by the $\epsilon$-expansion  in the (non-gauged) SU($N_f$)$\times$SU($N_f$) linear sigma model. 
We first review the argument of Pisarski and Wilczek \cite{Pisarski:1983ms} by examining
the structure of the fixed points of the renormalization group equations for  the quartic couplings  $\lambda_1$ and $\lambda_2$. 
We then examine the effective potential, following
Rudnick \cite{Rudnick:1978}, Iacobson and Amit \cite{Iacobson:1981jm, Amit:1978dk}, 
in order to see that  the transition can  actually be first-order. 

In $d=4-\epsilon$ dimensions,  the beta functions of the quartic couplings $\lambda_1$ and 
$\lambda_2$ at one-loop are obtained as follows:
\begin{eqnarray}
\label{eq:betafunc}
 \beta _1
 &=& - \epsilon \lambda_1+\frac{3}{8\pi ^2}\biggl(\frac{N_f^2 + 4}{3} \lambda_1^2 + \frac{4N_f}{3} \lambda_1\lambda_2 + \lambda_2^2 \biggr), \nonumber\\
\beta _2&=& - \epsilon \lambda_2+\frac{3}{8\pi ^2}\biggl( 2\lambda_1 \lambda_2 + \frac{2N_f}{3} \lambda_2^2 \biggr) . 
\end{eqnarray}
There are three kinds of the fixed point $(\lambda_1^\ast, \lambda_2^\ast)$: 
\begin{enumerate}
\item[(a)] $(\lambda_1^\ast, \lambda_2^\ast)=(0,0)$. 

\item[(b)]  $(\lambda_1^\ast, \lambda_2^\ast)=8\pi^2 \epsilon \left(\frac{1}{N_f^2 + 4},0\right)$.

\item[(c)]  $(\lambda_1^\ast, \lambda_2^\ast)
=8\pi^2 \epsilon\left( c, (1-6c)\frac{1}{2N_f}\right)$  \,($N_f \le \sqrt{3}$), \\
where $c$ is a solution of the equation: \\
$(N_f^4-8N_f^2+27)c^2+(N_f^2-9)c +3/4=0$.  
\end{enumerate}

The stability of these fixed points can be examined through the stability matrix defined by
$\hat \beta_{ij}\equiv  \partial \beta_i / \partial \lambda_j \, (i,j=1,2)$: 
\begin{equation}
\hat \beta 
= 
\left( 
\begin{array}{cc} 
   -\epsilon +\frac{\left[ 2 (N_f^2+4) \lambda_1^\ast  + 4N_f \lambda_2^\ast \right]}{8\pi ^2}
& \frac{\left[4 N_f  \lambda_1^\ast +6 \lambda_2^\ast\right]}{8\pi ^2} \\
\frac{\left[6 \lambda_2^\ast \right]}{8\pi ^2} 
& -\epsilon + \frac{\left[6 \lambda_1^\ast +4 N_f \lambda_2^\ast \right]}{8\pi ^2}\, 
\end{array} 
\right). 
\end{equation}
The fixed point (a), the Gaussian fixed point,  is IR-unstable. The fixed point (b) is IR-stable if 
$N_f < \sqrt{2}$.  This corresponds to the Wilson-Fisher fixed point  with 
$O(2N_f)$ critical exponent. When $N_f \ge \sqrt{2}$, it becomes IR-unstable in the $\lambda_2$ direction. The fixed point (c) exists if $N_f\leq \sqrt{3}$ and is IR-unstable if $N_f<\sqrt{2}$. 
Therefore, if $N_f > \sqrt{3}$ there is no IR-stable fixed points. 

 \begin{figure}
  \begin{center}
    \begin{tabular}{cc}
      \resizebox{38mm}{!}{\includegraphics{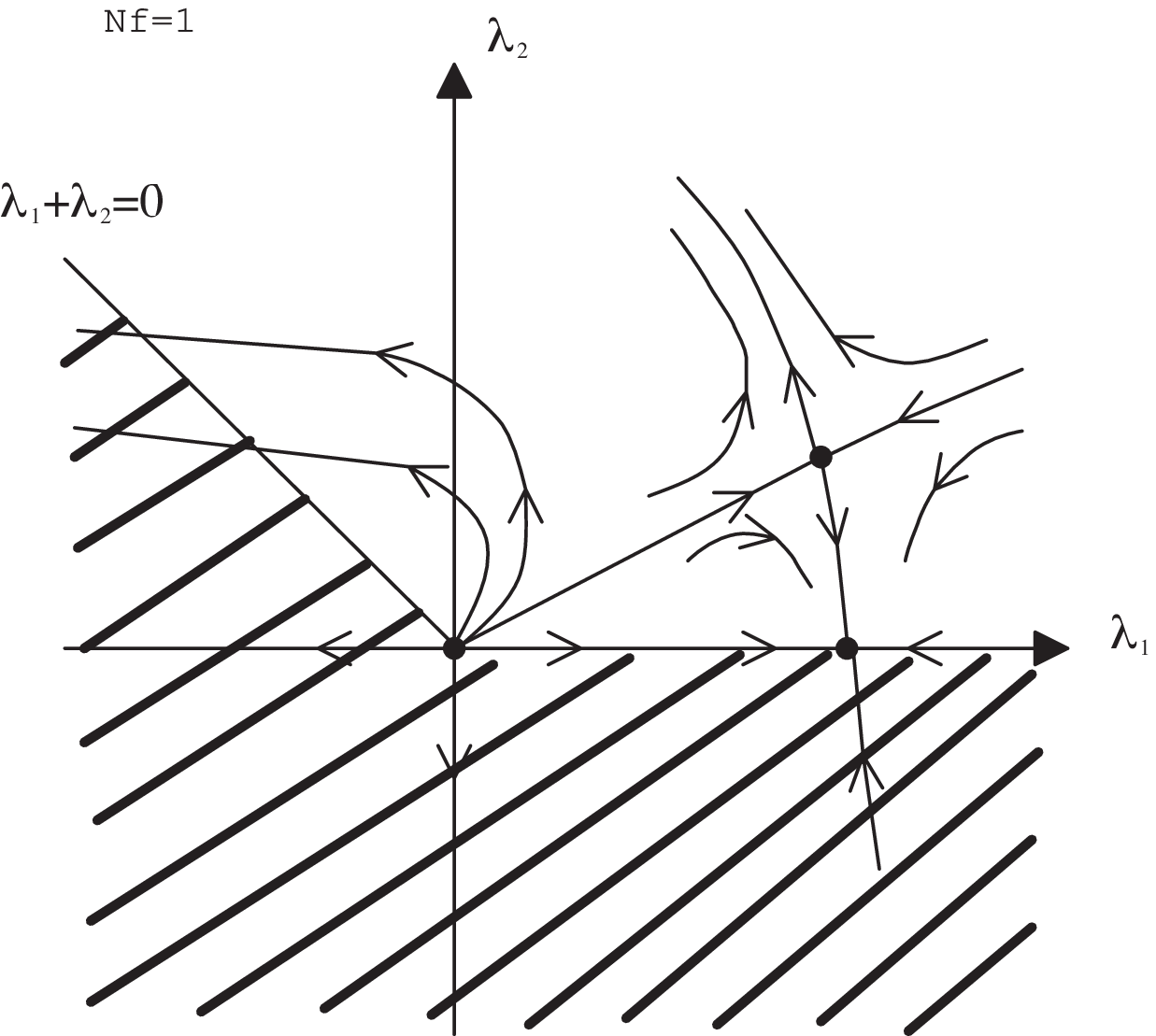}} &
      \resizebox{38mm}{!}{\includegraphics{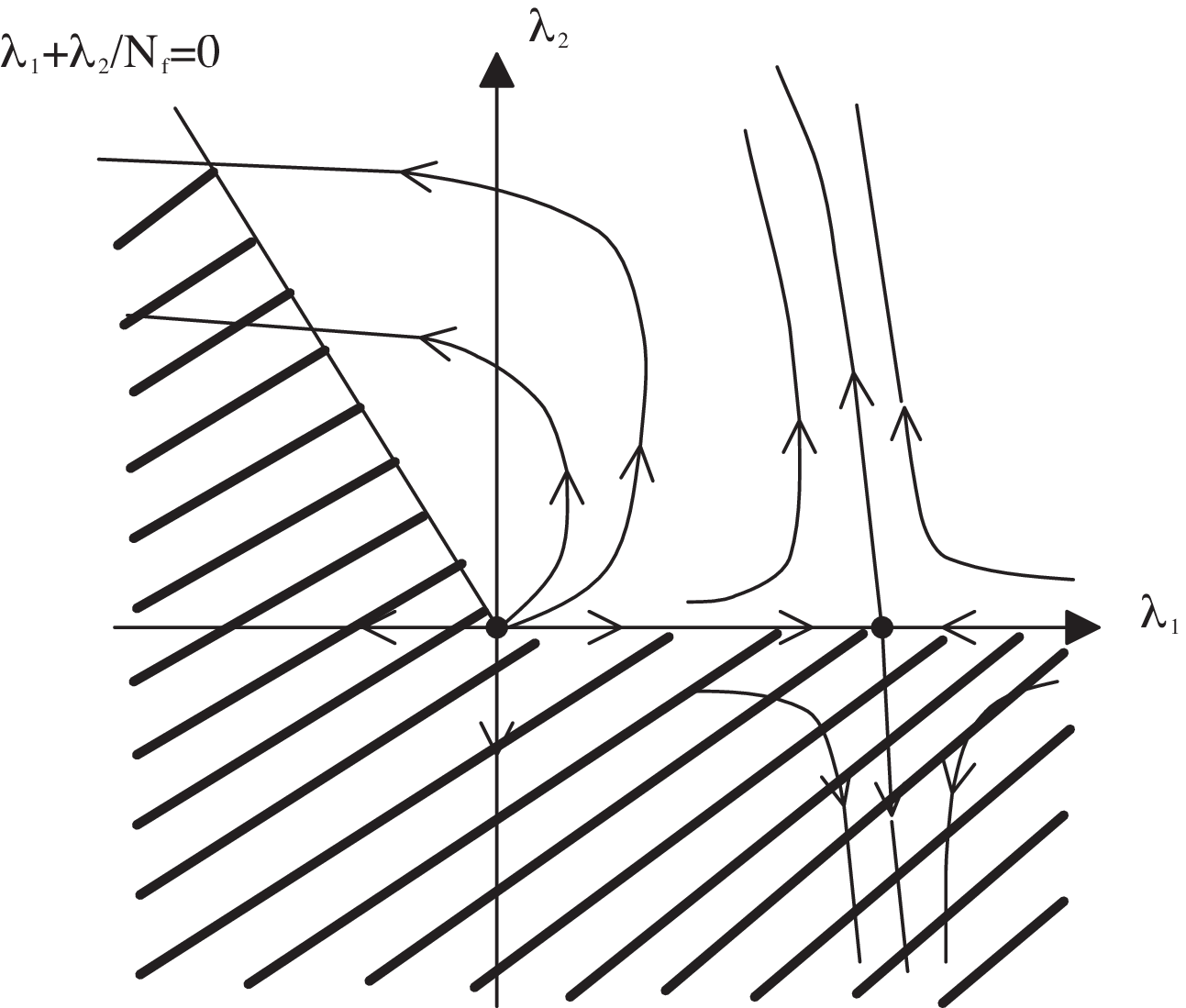}} \\
    \end{tabular}
  \end{center}
  \caption{The flow diagrams in the SU($N_f$)$\times$SU($N_f$) linear sigma model at $d=4-\epsilon$ 
for $N_f=1$ (left) and for $N_f > \sqrt{3}$ (right).
The arrows indicate the flows of the effective coupling constants toward IR direction.
In the shaded region, the effective potential Eq.~(\ref{eq:eff-potential-tree}) is unstable.}
   \label{fig:flow}
\end{figure}

The case of $N_f =2$ requires a special care. As discussed in the previous section, one can add the quadratic term Eq.~(\ref{eq:u1axial-breaking}) and take the O(4) limit. Then the quartic couplings reduce to the single coupling $\lambda_1+\lambda_2/2$ and the unstable direction no more exists.  The fixed point (b) is then  the IR-stable Wilson-Fisher fixed point  with O(4) critical exponent. 

 From these considerations,  
one can see that there is no IR-stable fixed points for  $N_f \ge {3}$.  This suggests
that the critical behavior of the restoration of chiral symmetry with $N_f \ge 3$  is not second order, 
although the possibility  is not completely excluded. 

In order to see that  the transition can actually  be first-order, 
we next examine the effective potential. 
At one-loop, the effective potential of the linear sigma model  is given by 
 \begin{eqnarray}
V_{\eff}(\phi)&=&\frac{1}{2}m_\Phi ^2\phi ^2
+\frac{1}{8}\biggl(\lambda_1 +\frac{\lambda_2}{N_f}\biggr)\phi ^4 \nonumber\\
&&+f\left[m_\Phi^2+\frac{3}{2}\left(\lambda_1+\frac{\lambda_2}{N_f}\right)\phi^2\right] 
\nonumber\\
&&+(N_f^2-1)f\left[m_\Phi^2+\frac{1}{2}\left(\lambda_1+3\frac{\lambda_2}{N_f}\right)\phi^2\right] 
\nonumber\\
&&+N_f^2f\left[m_\Phi^2+\frac{1}{2}\left(\lambda_1+\frac{\lambda_2}{N_f}\right)\phi^2\right], 
\end{eqnarray}   
where
\begin{equation}
f(x)=\frac{1}{64\pi ^2}x^2\biggl( \ln x-\frac{3}{2} \biggr). 
\end{equation}
For the consistency with the $\epsilon$-expansion, it is assumed that   
\begin{equation}
\lambda_1\sim\lambda_2 \sim \mathcal{O}(\epsilon) . 
\end{equation}

The first-order phase transition requires that the following conditions are fulfilled at some values of
the parameters:
\begin{equation}
\label{eq:first-order-condition-epsilon}
\frac{\partial V_{\eff}}{\partial \phi}\bigg|_{\phi=\phi_c}=0, \qquad V_{\eff}(\phi_c)=V_{\eff}(0). 
\end{equation}
\begin{figure}[t]
\begin{center}
\includegraphics[width=4.5cm]{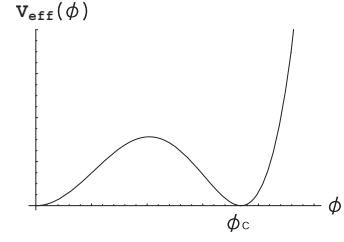}
\label{fig:1PT3d}
\caption{
Schematic form of the effective potential at the first-order phase transition.}
\end{center}
\end{figure}

\noindent
Now we consider the parameter region on "the stability boundary to $\mathcal{O}(\epsilon)$", 
following Rudnick \cite{Rudnick:1978}, Iacobson and Amit \cite{Iacobson:1981jm, Amit:1978dk}:
\begin{eqnarray}
\label{eq:stability-boundary}
\lambda_1+\lambda_2/N_f \sim \mathcal{O}(\epsilon^2) . 
 \end{eqnarray}
There, if one assumes
\begin{equation}
m_\Phi^2 \sim \mathcal{O}(\epsilon) ,\qquad \phi_c^2 \sim \mathcal{O}(\epsilon^{-1}), 
\end{equation}
the effective potential can be approximated as 
 \begin{eqnarray}
V_{\eff}(\phi) &=& \frac{1}{2}m_\Phi ^2\phi ^2
+\frac{1}{8}\biggl(\lambda_1 +\frac{\lambda_2}{N_f}\biggr)\phi ^4 \nonumber\\
&& +(N_f^2-1)f\left(\frac{\lambda_2}{N_f}\phi^2\right) + \mathcal{O}(\epsilon) . 
\label{eq:apxveff}
\end{eqnarray}
Then, by denoting 
\begin{equation}
m_\Phi^2 = 2 b \epsilon, \qquad 
\phi_c^2 = c \epsilon^{-1}  
\end{equation}
and 
\begin{equation}
\lambda_1+\frac{\lambda_2}{N_f} = 4 \delta \epsilon^2 , \qquad \lambda_2/N_f =  a \epsilon, 
\end{equation}
the conditions Eq.~(\ref{eq:first-order-condition-epsilon}) read explicity as 
\begin{eqnarray}
&& b c + \frac{1}{2} \delta c^2 + \frac{(N_f^2-1)}{64\pi^2} (ac)^2 [ \ln (ac)-3/2 ] = 0 , \\
&& b + \delta c +  \frac{(N_f^2-1)}{64\pi^2} (a^2 c) [ 2 \ln (ac)-2] = 0 .  
\end{eqnarray}
A solution to these conditions is given by
\begin{eqnarray}
b &=& \frac{(N_f^2-1)}{64 \pi^2} \, {\rm e}^{\left[1/2-\frac{32\pi^2}{(N_f^2-1)}\delta / a^2 \right]} \, 
a, \\
c &=& \frac{1}{a} \, {\rm e}^{\left[1/2-\frac{32\pi^2}{(N_f^2-1)}\delta / a^2 \right]} . 
\end{eqnarray}
Thus we can see that  
the restoration of the chiral symmetry with $N_f \ge 3$ can be first-order indeed 
in a parameter region close to the stability boundary, Eq.~(\ref{eq:stability-boundary}).

\section{Sphaleron\label{sec:sphaleron}}
In this section, we next examine the sphaleron energy \cite{Manton:1983nd,Klinkhamer:1984di} 
within the SU(2)$\times$U(1) gauged linear sigma model in order to clarify the criterion for 
the strength of the first-order  transition required for the electroweak baryogenesis.
For simplicity, we consider the one-family type and omit the U(1) hyper-charge interaction. 

When the vacuum preserves SU($N_f/2$)$_V$, we may assume the following anzatz
\begin{eqnarray}
&& A_i^a\frac{\sigma^a}{2}dx^i = -\frac{i}{g}f(\xi)dUU^{-1} ,  \\
&& \Phi = \frac{v}{\sqrt{2N_f}}h(\xi)\hat U,
\end{eqnarray}
\begin{gather}
U=\frac{1}{r}\begin{pmatrix} z & x+iy \\ -x+iy & z \end{pmatrix}, 
\quad 
\hat U= \left( \begin{array}{@{\,}cccc@{\,}}
  U &  &  & \\
   & U & & \\ 
   & & \ddots & \\ 
   & & & U 
\end{array} \right), 
\end{gather}
where
$\xi=gvr$ is the dimensionless radial coordinate,  $f(\xi)$ and $h(\xi)$ are the unknown functions which subject to the boundary conditions $f(0)=h(0)=0$ and $f(\infty)=h(\infty)=1$.  
If we introduce a $2\times2$ matrix variable $M$ as
\begin{equation}
M\equiv\frac{v}{\sqrt{2}}h(\xi)U,
\end{equation}
the energy functional of our model is written as
\begin{eqnarray}
E&=&\int d^3x \left\{ \frac{1}{4}F^a_{ij}F^a_{ij}+\frac{1}{2}\tr[(D_iM)^\dagger D_iM]\right. 
\nonumber \\
&&\left.-\frac{1}{4}\left(\lambda_1+\frac{\lambda_2}{N_f}\right)v^2\tr(M^\dagger M) \right. 
\nonumber \\
&&\left.+ \frac{1}{8} \left( \lambda_1+\frac{\lambda_2}{N_f} \right)(\tr M^\dagger M)^2 \right\} , 
\label{efsph}
\end{eqnarray}
where $v$ is the VEV determined through the effective potential at tree-level. 
We note that this result holds true even when one includes the explicit symmetry 
breaking terms as long as 
the last term of $\Delta {\cal L}_2$ and $\Delta {\cal L}_4$ vanishes identically, because 
the only effect of the symmetry breaking terms is then to shift the quadratic term of the effective potential.  

We now compare this result to the energy functional in the SM \cite{Manton:1983nd,Klinkhamer:1984di}:  
\begin{eqnarray}
E^{SM}&=&\int d^3x \left\{ \frac{1}{4}F^a_{ij}F^a_{ij}+
\frac{1}{2}\tr[(D_iM)^\dagger D_iM] \right.\nonumber \\
&& - \left.\frac{1}{2}\lambda v^2 \tr(M^\dagger M) 
+ \frac{1}{4}\lambda[\tr(M^\dagger M)]^2 \right\} . 
\end{eqnarray}
One can easily see that 
these energy functionals are equal to each other if one identifies the quartic couplings by the relation 
\begin{equation}
\frac{1}{2}\left(\lambda_1+\frac{\lambda_2}{N_f}\right)\leftrightarrow \lambda.  
\label{eq:replace1D}
\end{equation}
From this fact, one can easily estimate the sphaleron energy in our model. Namely, 
\begin{align}
E_{sph} &= E_{sph}^{SM} \notag \\
&= \left. \frac{4\pi v}{g}B\left(\frac{\lambda}{g^2}\right)
      \right\vert_{\lambda =\left(\lambda_1+\lambda_2/ N_f\right) / 2 } , 
\end{align}
where $B\left(\lambda/g^2\right)$ is the function given in 
\cite{Manton:1983nd,Klinkhamer:1984di}, whose numerical value is of order unity for 
all range of the variable $\lambda/g^2$.  Then the condition for the baryon asymmetry not to be washed out 
is given by 
\begin{equation}
\frac{v_c}{T_c}\gtrsim 1,  \label{eq:cri1F}
\end{equation}
just same as in the SM.  We assume this criterion in the following analysis.

\section{Analysis of finite temperature effective potential\label{sec:analysis-effective-potentail}}
In this section,  
we examine the critical behavior of the electroweak symmetry restoration at high temperature in 
the SU(2)$\times$U(1) gauged linear sigma model using
the finite temperature effective potential. To evaluate the effective potential, we adopt the ring-improved perturbation theory at one-loop. We first give some analytical results in the high temperature expansion. 
We next examine the effective potential numerically and estimate semi-quantitatively the strength of the first-order phase transition.

\subsection{Finite temperature effective potential of \\
SU(2)$\times$U(1) gauged linear sigma model}

In the one-loop approximation with the ring-improvement, 
the effective potential of the SU(2)$\times$U(1) gauged linear sigma model
may be written as
\begin{eqnarray}
V(\phi) &=&V_0(\phi) +V_1^{(0)}(\phi) + \nonumber\\
&& V_1^{(T)}(\phi,T)+ V_{\ring}(\phi,T), 
\end{eqnarray}
where $V_0$ is the tree-level effective potential, $V_1^{(0)}$ and $V_1^{(T)}$ are the one-loop contributions at zero and finite temperature, respectively, and 
$V_{\ring}(\phi,T)$ is the contribution from the ring diagrams.
$V_0$,  the tree-level effective potential, is given by Eq.~(\ref{eq:eff-potential-tree}). 
When one includes the explicit symmetry breaking terms, Eq.~(\ref{eq:symmetry-breaking}), 
the coefficient of the quadratic term should be shifted as $m_{\Phi}^2 \to m_{\Phi}^2+\Delta m^2-c$.

$V_1^{(0)}$, the one-loop contribution at zero temperature, is given by 
\begin{align}
V_1^{(0)}(\phi) =&\frac{1}{64\pi^2}\sum_{i}n_im_i^4(\phi)
             \left[\ln\frac{m_i^2(\phi)}{\mu^2}-C_i\right] \label{eq:1loop0} , 
\end{align}
where $i$ runs over all scalar and  vector bosons.  $n_i$ and $m_i(\phi)$  are the number of degrees of freedom and the effective  masses depending on $\phi$, respectively, which are  given in 
TABLE~\ref{tab:lsmm}--\ref{tab:lsm+sb}. 
$C_i$ are the constants given by $C_i=3/2$ for scalar bosons and $C_i=5/6$ for gauge bosons.
We have chosen the Landau gauge and have used the $\overline{\rm MS}$ scheme to renormalize the ultraviolet divergences at the renormalization scale $\mu$. At one-loop, the VEV is determined by 
\begin{equation}
 \frac{\partial [V_0(\phi)+V_1^{(0)}(\phi) ]}{\partial \phi} \bigg|_{\phi=\phi_0} = 0. 
\end{equation}

\begin{figure}[t]
\begin{center}
\includegraphics[width=8cm]{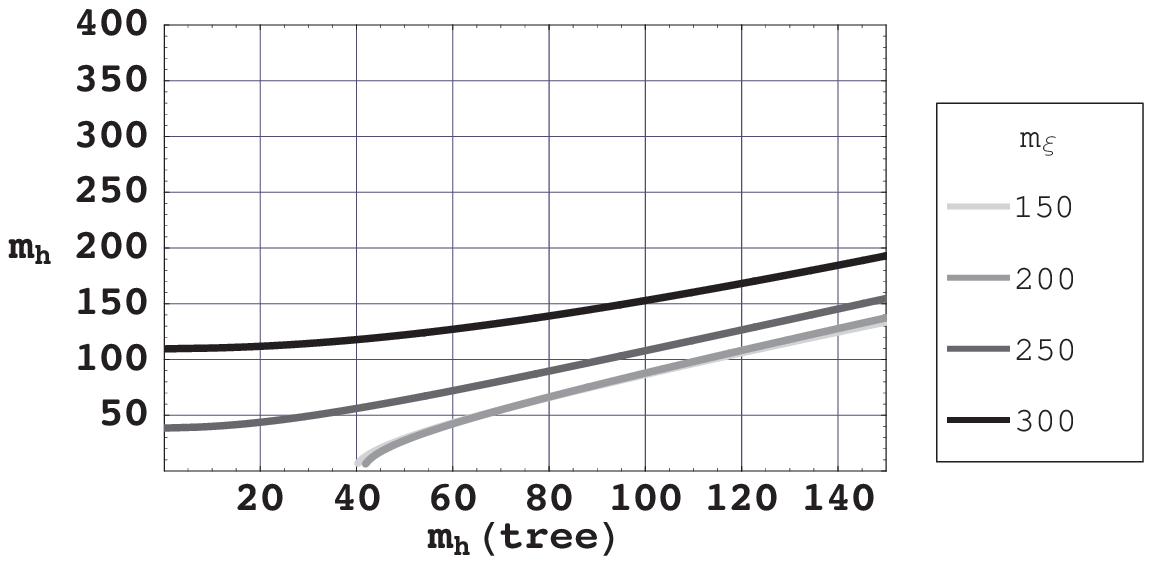}
\caption{$m_h$  as a function of 
$(\lambda_1+\lambda_2 /N_f)^{1/2} \phi_0$ for the case $N_f = 4$. }
\label{mh-N4}
\includegraphics[width=8cm]{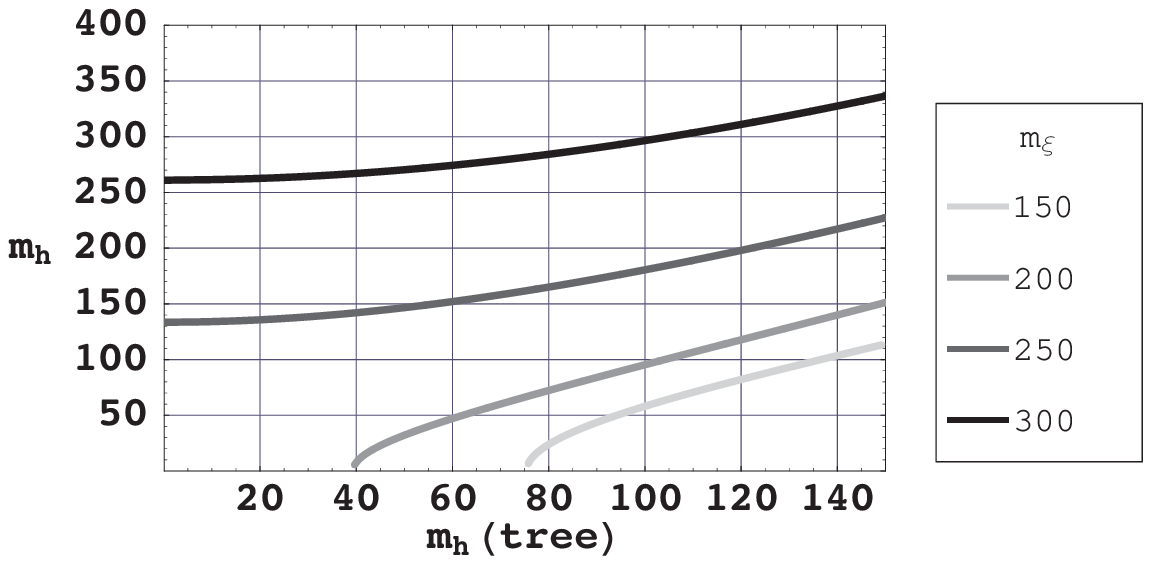}
\caption{$m_h$  as a function of 
$(\lambda_1+\lambda_2 /N_f)^{1/2} \phi_0$ for the case $N_f = 8$. }
\label{mh-N8}
\end{center}
\end{figure}

At one-loop, the mass of the neutral scalar field, $m_h$, acquires a rather large correction, the size of which depends on $N_f$ and $\lambda_2/N_f$ as well as $\lambda_1+\lambda_2/N_f$.  
Then, we adopt the following definition for the (renormalized) mass of the neutral scalar field, $m_h$:
\begin{equation}
\label{eq:mass-h}
m_h^2\equiv\frac{\partial^2 [V_0(\phi)+V_1^{(0)}(\phi)]}{\partial\phi^2}\bigg|_{\phi=\phi_0}.
\end{equation}
As for the mass of the extra scalar fields, $m_\xi$, we adopt the formula at  the tree level:
\begin{equation}
\label{eq:mass-xi}
m_\xi^2 \equiv \frac{\lambda_2}{N_f} \, \phi_0^2 . 
\end{equation}
FIG.~\ref{mh-N4} shows 
the plot of $m_h$ as a function of $[m_h]_{\rm tree} =(\lambda_1+\lambda_2/N_f)^{1/2} \phi_0$
for several values of $m_\xi$ and $N_f=4$. FIG.~\ref{mh-N8} shows a similar plot for $N_f=8$. 

The one-loop contribution at finite temperature, $V_1^{(T)}$, is given by 
\begin{align}
V_1^{(T)}(\phi,T) =\frac{T^4}{2\pi^2}\sum_{i}n_iJ_B[m_i^2(\phi)/T^2] , 
\end{align} 
where $i$ runs over all scalar  and vector bosons. $J_B$ is defined by  
\begin{equation}
J_B(a)=\int_{0}^{\infty}dx~x^2 \ln\left(1-e^{-\sqrt{x^2+a}}\right).
\end{equation}
In the high temperature limit where $m(\phi) \ll T$,  it can be expanded as follows:
\begin{align}
J_B(m^2/T^2)= 
&-\frac{\pi^4}{45}+\frac{\pi^2}{12}\frac{m^2}{T^2}-\frac{\pi}{6}\left(\frac{m^2}{T^2}\right)^{3/2} \notag \\
&-\frac{1}{32}\frac{m^4}{T^4}\ln\frac{m^2}{a_bT^2}+\mathcal{O}\left(\frac{m^6}{T^6}\right) , 
\end{align}
where $a_b=16\pi^2\exp(3/2-2\gamma_E)$($\ln a_b\approx 5.4076$).
 
In the ordinary perturbation theory at finite temperature, the perturbative expansion breaks down near the critical temperature due to the exitence of 
the higher-loop IR divergent diagrams in the massless limit.
This problem can be partially avoided by resumming the contributions from the ring diagrams 
which are the most dominant IR contributions at each order of the perturbative expansion, 
though the coupling constants must satisfy certain constraints to suppress the other higher-loop 
contributions \cite{Fendley:1987ef,Espinosa:1992gq,Parwani:1991gq,Anderson:1991zb,
Carrington:1991hz, Dine:1992wr,Arnold:1992rz,Fodor:1994bs}.

\begin{figure}
  \centering
  \includegraphics[width=160pt,clip]{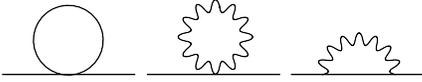}
  \caption{The one-loop self energy diagrams of scalar bosons. First one comes from the self-interactoin, second and third one are contributions from the SU(2)$\times$U(1) gauge bosons. In the one-family type with $N_f=8$, there are also the contributions from the gluon.}
  \label{fig:1D-loop}
\end{figure}

\begin{figure}
  \centering
  \includegraphics[width=160pt,clip]{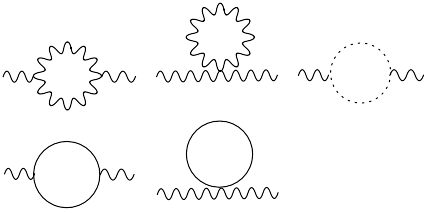}
  \caption{The one-loop vacuum polarization diagrams of SU(2)$\times$U(1) gauge bosons. First and second one comes from the self interactions, 
   third one is the contribution from the ghost, fourth and fifth one are the contribution from the scalar bosons.}
  \label{fig:gloop}
\end{figure}

One can include the contribution of ring diagrams,  $V_{\ring}(\phi,T)$,  by replacing $m_i^2(\phi)$ in $V_1^{(0)}$ and $V_1^{(T)}$
with the effective T-dependent masses $\mathcal{M}_i^2(\phi,T)\equiv m_i^2(\phi)+\Pi_i$,
where $\Pi_i$ is the self-energy of a particle $i$ in the IR limit where the Matsubara frequency and the momentum of the external fields go to zero 
and in the leading order of $m_i(\phi)/T$.
For the scalar bosons, there are the contributions from the diagrams shown in FIG.~\ref{fig:1D-loop}.
For the longitudinal components of the gauge bosons, there are the contributions from the diagram 
shown in FIG.~\ref{fig:gloop}, 
while the transverse components do not receive the correction. 
The explicit expressions of the effective T-dependent masses depend on how the gauge interactions are introduced in the sigma model  and are given in appendix.

After all, the one-loop ring-improved effective potential is given by
\begin{widetext}
\begin{align}
V(\phi) =&V_0(\phi) +V_1^{(0)}(\phi) +V_1^{(T)}(\phi,T)+ V_{\ring}(\phi,T)  \notag \\
             =&V_0(\phi)+\sum_{i}n_i\left[\frac{1}{64\pi^2}\mathcal{M}_i^4(\phi,T)
             \left\{\ln\frac{\mathcal{M}_i^2(\phi,T)}{\mu^2}-C_i\right\}
                +\frac{T^4}{2\pi^2}J_B[\mathcal{M}_i^2(\phi,T)\beta^2]\right].
\end{align}
\end{widetext}

 A comment on the validity of the ring-improved perturbation theory is in order. For simplicity, 
let us consider
only the contribution of the scalar fields,  neglecting the contribution from the gauge fields.
This approximation is expected to be valid as long as the scalar fields are heavier than the gauge fields. 
By inspecting the higher order diagrams for the scalar field self-energies, one can see that 
the non-ring diagrams are suppressed with respect to the ring diagrams at least 
by the following factors in the symmetric phase, 
\begin{eqnarray}
\label{eq:beta-ring}
\beta_{\lambda_1+\lambda_2/N_f} 
&\equiv& \frac{1}{4\pi}\left(\lambda_1+\frac{\lambda_2}{N_f}\right)\frac{T}{m_{\rm eff}} , \notag\\
\beta_{\lambda_2} 
&\equiv& \frac{N_f^2}{4\pi}\left(\frac{\lambda_2}{N_f}\right)\frac{T}{m_{\rm eff}} , 
\end{eqnarray}
where 
\begin{equation}
m_{\rm eff}^2 \equiv m_\Phi^2+\frac{T^2}{12}\sum_in_ia_i = m_\Phi^2+\frac{b}{12}T^2.  
\end{equation}
On the other hand, in the broken phase, the suppression factors depend on various mass parameters. 
Among them, the largest factors in the large $N_f$ limit are given by 
the loops of the neutral scalar boson $h$ and the NG bosons $\pi$ and $\eta$,   
\begin{eqnarray}
\label{eq:beta-ring-broken-phase}
\bar \beta_{\lambda_1+\lambda_2/N_f} 
&\equiv& \frac{N_f^2}{4\pi}\left(\lambda_1+\frac{\lambda_2}{N_f}\right)
\frac{T}{(m_{\rm eff}^2+a_h \phi^2)^{1/2}} , \notag\\
\bar \beta_{\lambda_2} 
&\equiv& \frac{N_f^2}{4\pi}\left(\frac{\lambda_2}{N_f}\right)
\frac{T}{(m_{\rm eff}^2+a_\pi \phi^2)^{1/2}} . 
\end{eqnarray}
Therefore, in order to guarantee the validity of the ring-improved perturbation theory, it is required that $\beta_{\lambda_1+\lambda_2/N_f}, \bar \beta_{\lambda_1+\lambda_2/N_f}\ll 1$ 
and $\beta_{\lambda_2}, \bar\beta_{\lambda_2}\ll 1$, while  
the perturbative expansion at zero temperature is valid  for $N_f^2(\lambda_1+\lambda_2/N_f)/(4\pi)^2\ll 1$ and $N_f\lambda_2/(4\pi)^2 \ll 1$. 
In the following analysis of the effective potential, we will examine whether these conditions are satisfied near the critical temperature. 

\subsection{Analytical results in the high temperature expansion}

We first examine the effective potential analytically using the high temperature expansion in order to  get some insight how $\phi_c / T_c$ depends on  the number of flavor $N_f$ and the quartic couplings  $\lambda_1+\lambda_2/N_f$ , $\lambda_2/N_f$.
For simplicity, we consider only the contribution of the scalar fields as discussed before. 
In the high temperature limit,  the one-loop effective potential can be expanded as 
\begin{eqnarray}
\label{eq:V-in-HTE}
V(\phi,T)&\simeq& \frac{1}{2}m_{\rm eff}^2\phi^2
-\frac{T}{12\pi}\sum_in_i  \left[ ( {\cal M}^2_i )^{3/2} - ( m_{\rm eff}^2)^{3/2}\right] \nonumber \\
&&+\frac{1}{8}\left(\lambda_1+\frac{\lambda_2}{N_f}\right)\phi^4  +{\cal O}({\cal M}^6/T^6), 
\label{eq:veff-hte}
\end{eqnarray}
where ${\cal M}^2_i = m_{\rm eff}^2 + a_i \phi^2$.  
We have subtracted the terms which do not depend on $\phi$, 
\begin{equation}
V(0,T)= \sum_i n_i \left\{ - \frac{\pi^2 T^4}{90} + \frac{T^2}{24} m_{\rm eff}^2  - \frac{T}{12\pi} 
(m_{\rm eff}^2)^{3/2} \right\}, 
\end{equation}
and  have neglected the term $-\frac{\sum_i n_i}{64\pi^2} {\cal M}_i^4 \ln (a_b T^2/\mu^2)$ 
which may be regarded to be small compared to the tree-level terms. 
In order that the effective potential is real for all values of $\phi$, it must be satisfied that 
$m_{\rm eff}^2=m_\Phi^2+\frac{b}{12}T^2\geq 0$, namely, 
\begin{equation}
T^2\geq -12m_\Phi^2/b\equiv T_1^2.
\end{equation}
(At the tree level, $T_1^2=6(\lambda_1+\lambda_2/N_f) \phi_0^2 / b$.)

At the temperature $T_1$, $m_{\rm eff}^2$ is equal to zero and ${\cal M}^2_i = a_i \phi^2$. Then, it follows
\begin{eqnarray}
V(\phi,T_1)\simeq -\frac{T_1}{12\pi}\sum_in_ia_i^{3/2}\phi^3+\frac{1}{8}\left(\lambda_1+\frac{\lambda_2}{N_f}\right)\phi^4. 
\end{eqnarray}
At this temperature, the origin of the effective potential is unstable because the $\phi^3$ term  is negative and there is a minimum point of the potential 
where the scalar field has a nonzero VEV, 
\begin{eqnarray}
\label{eq:phi-1}
\phi_1=\frac{T_1}{2\pi}\frac{\sum_in_ia_i^{3/2}}{\lambda_1+\lambda_2/N_f}.
\end{eqnarray}
Above the temperature $T_1$, $m_{\rm eff}^2$ becomes positive, and the minimum point appears at the origin other than $\phi_1$. This signals the first-order phase transition, and one may expect a first-order phase transition at the temperature $T_c \, (>T_1)$ where 
the effective potential satisfies the conditions, 
\begin{equation}
\label{eq:first-order-condition}
\frac{\partial V}{\partial \phi}(\phi_c,T_c)=0 ,  \qquad
V(\phi_c,T_c)=V(0,T_c) .   
\end{equation}

One may consider the ratio $\phi_1/T_1$ as an estimate of the strength of the first-order transition and
Eq.~(\ref{eq:phi-1}) suggests that the first-order transition becomes stronger as one approaches the stability boundary, $(\lambda_1+\lambda_2/N_f) \simeq 0$.  
For a large $N_f$ and $ 0 \lesssim (\lambda_1+\lambda_2/N_f) \ll  \lambda_2/N_f \lesssim 16\pi^2/N_f^2$,  one obtains  
\begin{eqnarray}
\label{eq:phi-1-stability-boundary}
\frac{\phi_1}{T_1}  
&\simeq& 2 N_f  \left( \frac{\lambda_2/N_f}{\lambda_1 + \lambda_2/N_f} \right)
\left[ \frac{N_f^2}{16\pi^2} \frac{\lambda_2}{N_f} \right]^{1/2}  \nonumber\\
&=&
2 N_f \left[ \frac{m_\xi^2}{m_h^2} \right]_{\rm tree} \left[ \frac{N_f^2}{16\pi^2} \frac{\lambda_2}{N_f} \right]^{1/2} , 
\end{eqnarray}
where we have used the mass relations at the tree level, $m_h^2 = (\lambda_1+\lambda_2/N_f) \phi_0^2$ and $m_\xi^2 = (\lambda_2/N_f) \phi_0^2$.  
From Eq.~(\ref{eq:phi-1-stability-boundary}), we can see  that $\phi_1/T_1$ is enhanced   
for a large $N_f$ and for a large mass ratio $m_\xi^2/m_h^2$. 
We note in this case that 
the ring-improved perturbation theory is valid because one has
\begin{eqnarray}
\bar \beta_{\lambda_1+\lambda_2/N_f} &\simeq& 
\left(\frac{\lambda_1+\lambda_2/N_f}{\lambda_2/N_f}\right)^{3/2} \ll 1 , \nonumber \\
\bar \beta_{\lambda_2}&\simeq& \left(\frac{\lambda_1+\lambda_2/N_f}{\lambda_2/N_f}\right)^{1/2} \ll 1 . 
\end{eqnarray}
Thus,  the parameter region close to the stability boundary can be accessible within the ring-improved perturebation theory. 
On the other hand, 
the high temperature expansion is valid as long as  
\begin{equation}
\label{eq:HTE-validity-1}
\frac{\phi_1}{T_1} \, \, \lesssim \, \, \frac{1}{a_\xi^{1/2}}  = \left(\frac{1}{\lambda_2/N_f} \right)^{1/2} 
= \frac{ v  }{m_\xi } . 
\end{equation}
Therefore, in the high temperature expansion, 
the strong first-order transition around  the stability boundary can be examined  
only when $m_\xi$ is small compared to the weak scale $v$ ($=\phi_0$ for the one-family type). 

Another interesting parameter region one may consider is the region where the first-order transition turns into a second order transition or a crossover behavior. 
Such region can be characterized by the condition: $\phi_c/T_c \rightarrow 0$.  
In order to locate such parameter region,  
we expand the effective potential in terms of $\phi^2/T^2$ as 
\begin{equation}
V(\phi,T)
=T^4\left[c_1\frac{\phi^2}{T^2}+c_2\frac{\phi^4}{T^4}+c_3\frac{\phi^6}{T^6}
+{\cal O}\left(\frac{\phi^8}{T^8}\right)\right],
\label{pot_p6}
\end{equation}
where the coefficients are given within the high temperature expansion by
\begin{eqnarray}
c_1&=&\frac{1}{2}\frac{m_{\rm eff}^2}{T^2} \left(1-\frac{1}{4\pi}\sum_in_ia_i \frac{T}{m_{\rm eff}}\right) ,  \\
c_2&=&\frac{1}{8}\left(\lambda_1+\frac{\lambda_2}{N_f}-\frac{1}{4\pi}\sum_in_ia_i^2 \frac{T}{m_{\rm eff}}\right) ,  \\
c_3&=&\frac{1}{192\pi}\sum_in_ia_i^3 \left(\frac{T}{m_{\rm eff}}\right)^3 . 
\label{calv2}
\end{eqnarray}
Then  Eq.~(\ref{eq:first-order-condition})  is solved by
\begin{eqnarray}
\frac{\phi_c^2}{T_c^2}=\sqrt{\frac{c_1}{c_3}}=\frac{-c_2}{2c_3}  
\end{eqnarray}
for $c_1 \ge 0$, $c_2 \le 0$ and $c_3 > 0$ and the condition $\phi_c/T_c \rightarrow 0$ implies that $c_1= c_2 = 0$.  
This gives the critical end line: 
\begin{eqnarray}
\label{eq:critical-end-line}
\lambda_1+\frac{\lambda_2}{N_f}=\left|2+\frac{2}{N_f^2-2}\right|^{\frac{1}{2}}\frac{\lambda_2}{N_f}, 
\label{mh0}
\end{eqnarray}
or
\begin{equation}
\label{eq:critical-end-line-mass}
\left[ \frac{m_h^2}{m_\xi^2} \right]_{\rm tree} = \left|2+\frac{2}{N_f^2-2}\right|^{\frac{1}{2}} . 
\end{equation}
In this case, 
$c_1=0 $ implies that $m_{\rm eff}/T = \sum_i n_i a_i / 4\pi$. 
Then, 
we note that 
\begin{eqnarray}
\beta_{\lambda_1+\lambda_2/N_f} 
&\simeq& 
\frac{1}{N_f^2} \frac{\left(\lambda_1+\lambda_2/N_f\right)}
{\left(\lambda_1+\lambda_2/N_f\right) +\left(\lambda_2/N_f\right)} , \\
\beta_{\lambda_2} 
&\simeq& 
\frac{\left(\lambda_2/N_f\right)}
{\left(\lambda_1+\lambda_2/N_f\right) +\left(\lambda_2/N_f\right)} . 
\end{eqnarray}
Therefore, the ring improved perturbation theory is valid for 
$(\lambda_1+\lambda_2 / N_f) \gtrsim \lambda_2 / N_f (>0) $. 
On the other hand, 
the high temperature expansion is valid for 
\begin{equation}
 \frac{N_f^2}{4\pi} \left[ \left(\lambda_1+\lambda_2/N_f\right) +\left(\lambda_2/N_f\right)\right] \lesssim 1. 
\end{equation}
From Eq.~(\ref{eq:critical-end-line-mass}),  we can see that  when the mass of the neutral scalar field becomes large and exceeds those of the extra scalar fields the critical behavior becomes 
the second order transition, or a crossover behavior.  
(Within the above anaysis, if the mass ratio $[m_h^2/m_\xi^2]_{\rm tree}$ becomes larger 
than the critical value,
the transition is second order because  $c_2$ is always positive when $c_1=0$. )
 
These analytical results obtained within the high temperature expansion suggest that 
as long as $m_h \ll   m_\xi \, (< v)$,  one would not encounter the critical end line and 
the first-order transition can remain strong.
Then, the question is how large $m_\xi$ and $m_h$ can be, while keeping the strongly first-order transition with $\phi_c/T_c \simeq {\cal O}(1)$.  To examine this point, one should go beyond
the high temperature expansion. 

\subsection{Numerical results}

We next examine the effective potential numerically.  
The condition Eq.~(\ref{eq:first-order-condition}) is solved
for various parameters,  $m_h$, $m_\xi$ and 
$N_f$  (see Eqs.~(\ref{eq:mass-h}), (\ref{eq:mass-xi}) for definition), 
and  $\phi_c/T_c$ is evaluated in order to estimate the strength of the first-order phase transition.  
The renormalization scale is set to the electroweak scale as $\mu=v$.  
The mass parameter $m_\Phi^2$ is chosen so that  at zero temperature the electroweak symmetry is spontaneously broken at $\phi_0 = v$ for the one-family type and at $\phi_0=\sqrt{(N_f/2)} \, v$ for 
the partially gauged type.  

\subsubsection{SU($N_f$)$\times$SU($N_f$) linear sigma model} 
We first  show the  numerical result for 
the non-gauged SU($N_f$)$\times$SU($N_f$) linear sigma model 
in the one-family type ($\phi_0 = v$). 

\begin{figure}[t]
\begin{center}
\includegraphics[width=7cm]{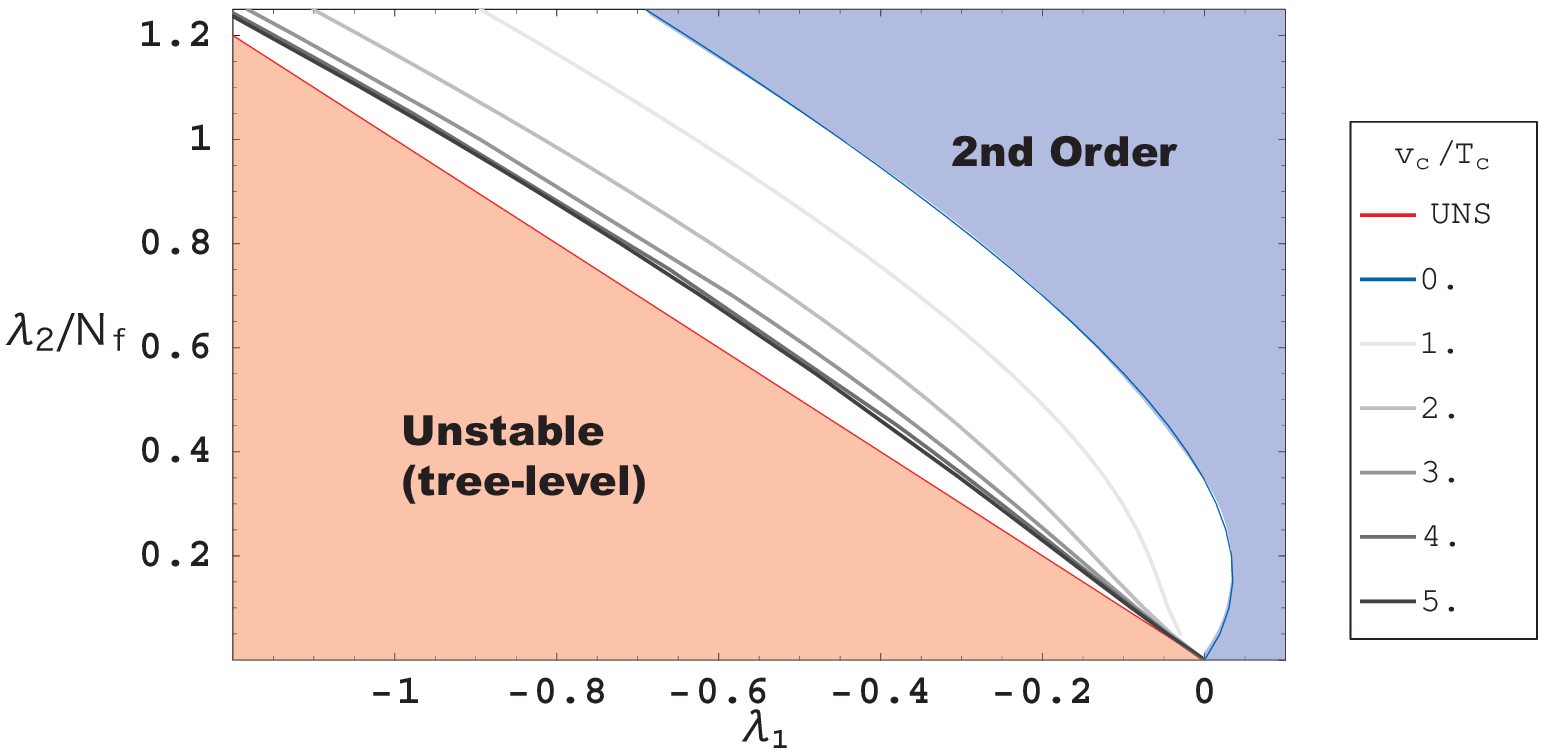}
\caption{(Color on line) 
$\lambda_1$--$\lambda_2/N_{f}$ diagram of 
the critical behavior of the chiral symmetry restoration in (non-gauged) 
SU($N_f$)$\times$SU($N_f$) linear sigma model with $N_f=4$.
The gray lines show the contours of  $\phi_c/T_c$. 
Above $\phi_c/T_c \simeq 0$ (the blue line), the phase transition becomes a second order. 
}
\label{PD-1F-N4}
\vspace{1em}
\includegraphics[width=7cm]{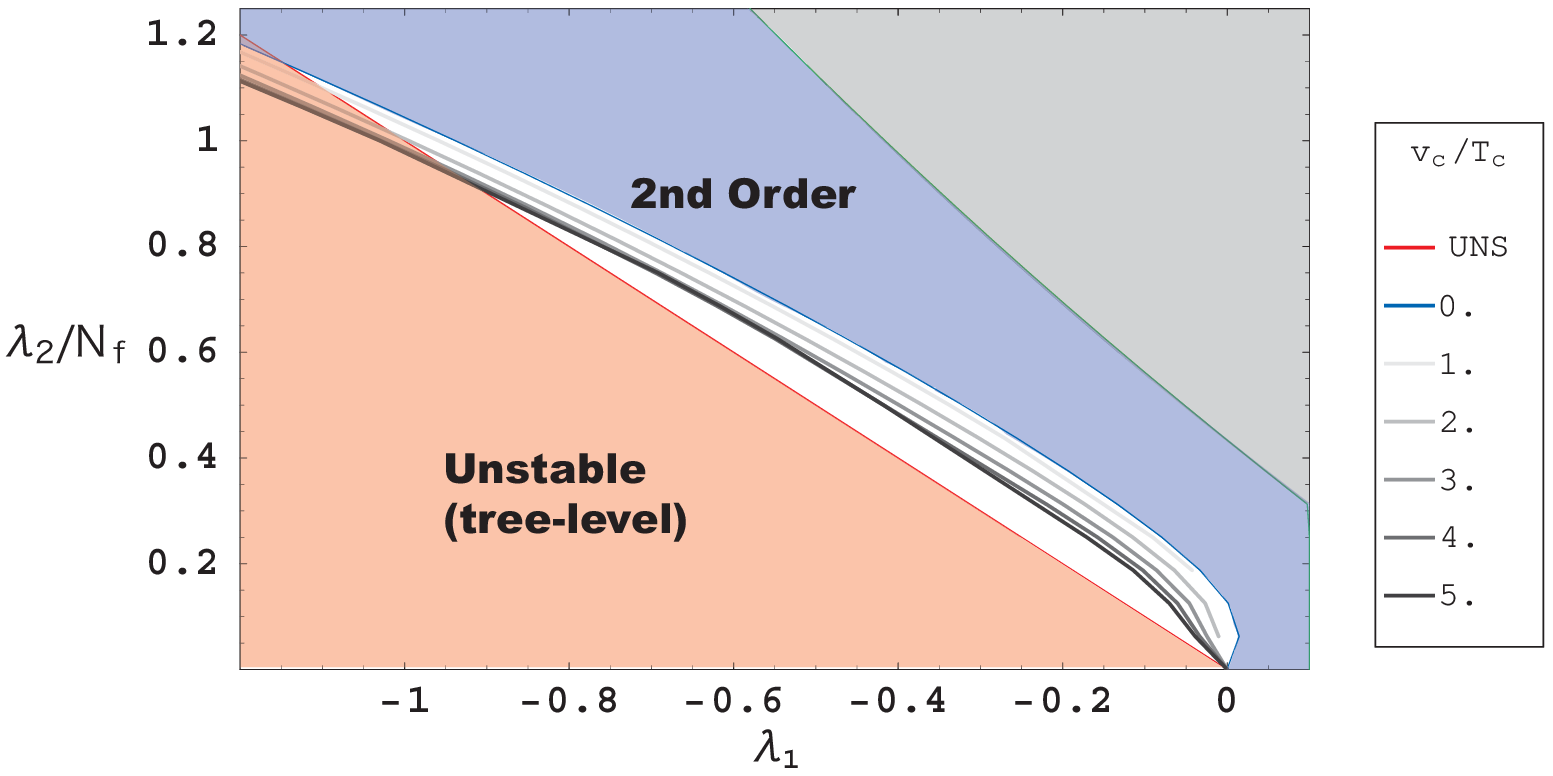}
\caption{(Color on line)
$\lambda_1$--$\lambda_2/N_{f}$ diagram of 
the critical behavior of the chiral symmetry restoration in (non-gauged) 
SU($N_f$)$\times$SU($N_f$) linear sigma model with $N_f=8$.
}
\label{PD-1F-N8}
\end{center}
\end{figure}

FIG.~\ref{PD-1F-N4} is  a diagram which shows the coupling dependence of 
the critical behavior of chiral symmetry restoration 
on the $\lambda_1$--$\lambda_2/N_f$ plane for $N_f = 4$.  We clearly see that the region of the strongly first-order transition lies
above the stability boundary $\lambda_1+\lambda_2/N_f = 0$, where      
the contours of $\phi_c/T_c$ run in almost parallel with the boundary for a large $\lambda_2/N_f$, 
$\lambda_2/N_f \gtrsim 0.7$.  
The critical end line is located around $\lambda_1+\lambda_2/N_f \simeq  0.5$ and 
the transition becomes second order  when $\lambda_1+\lambda_2/N_f$ gets large further. 
FIG.~\ref{PD-1F-N8} is a similar diagram for $N_f = 8$. 
%

\begin{figure}[t]
\begin{center}
\includegraphics[width=7cm]{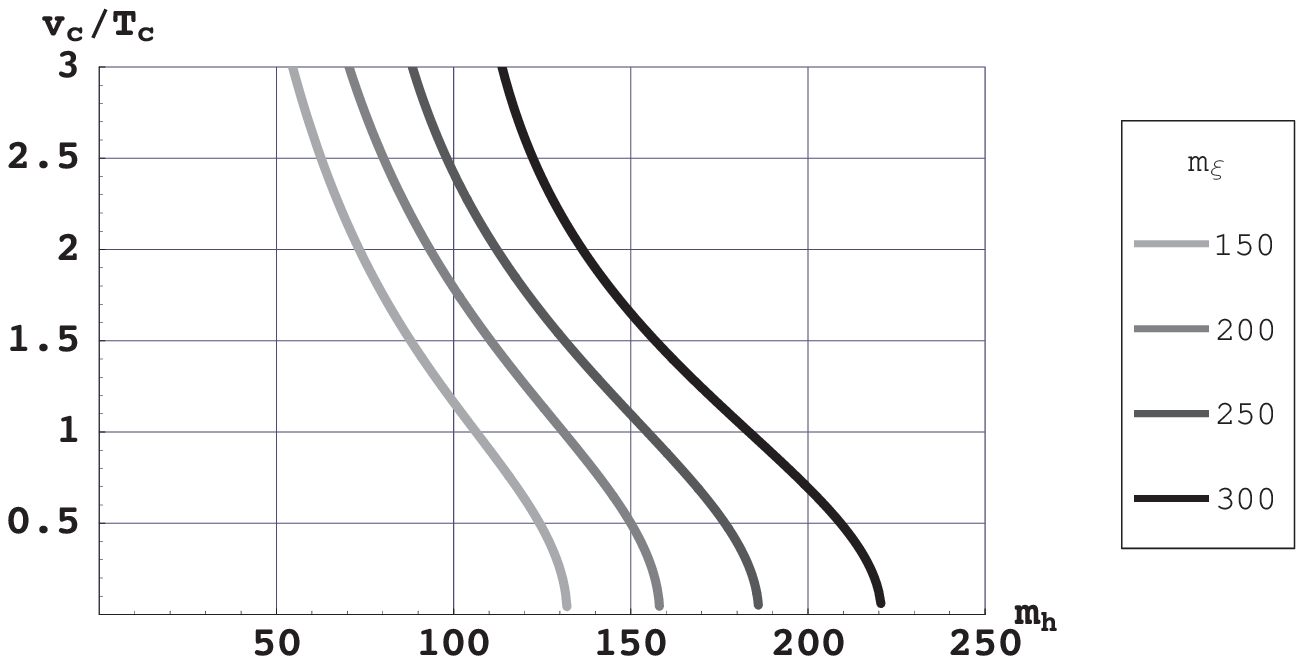}
\caption{$\phi_c/T_c$ as a function of $m_h$ for $N_f=4$. 
}
\label{mxi-1F-N4}
\vspace{1em}
\includegraphics[width=7cm]{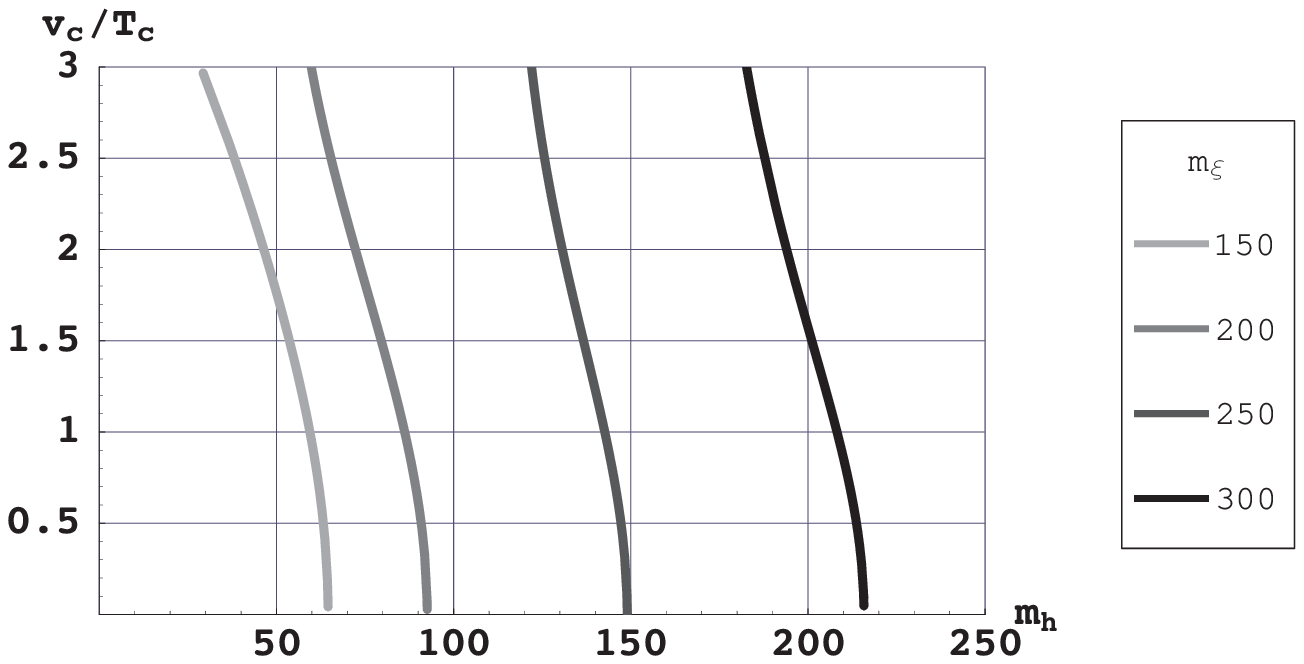}
\caption{
$\phi_c/T_c$ as a function of $m_h$ for $N_f=8$. 
}
\label{mxi-1F-N8}
\vspace{1em}
\includegraphics[width=7cm]{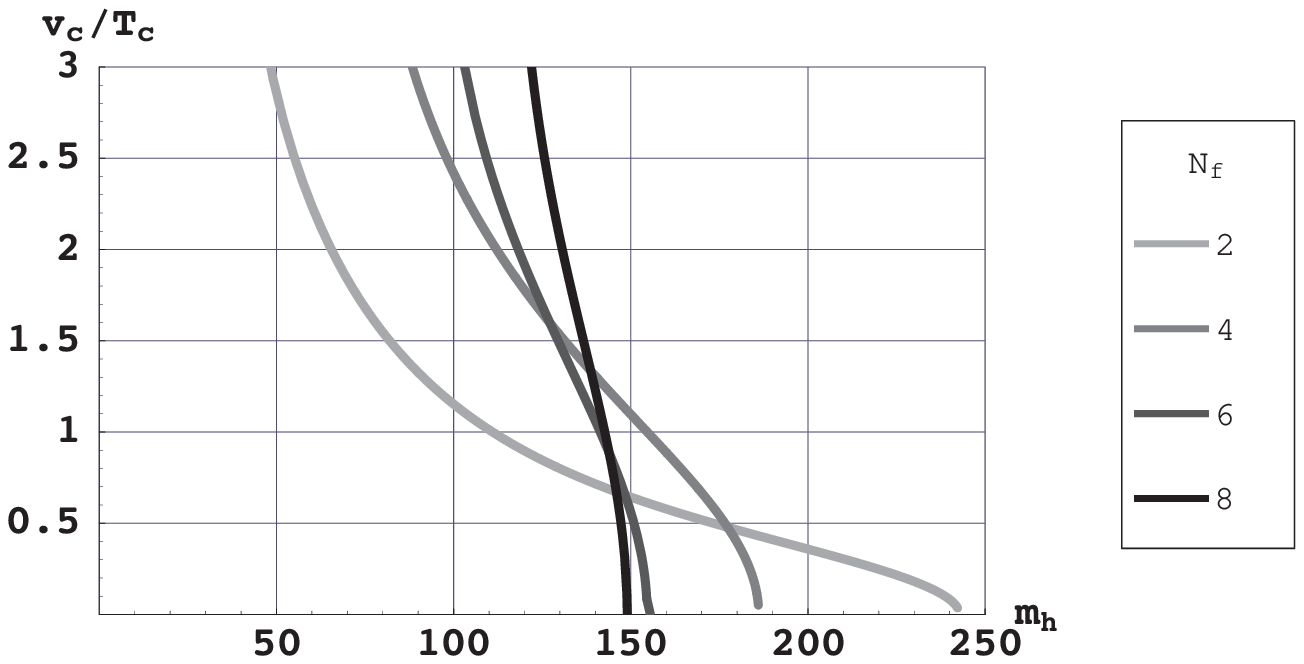}
\caption{
$N_f$ dependence of $\phi_c/T_c$ with $m_{\xi}=250$ GeV fixed.}
\label{Nf-1F-mxi250}
\end{center}
\end{figure}

In FIG.~\ref{mxi-1F-N4},  $\phi_c/T_c$ is plotted as a function of $m_h$ for several values of $m_\xi$
and for $N_f = 4$. The unit of mass is in GeV. 
We can see that with $m_\xi$ fixed, $\phi_c/T_c$ decreases  and 
finally vanishes identically as $m_h$ increases.  On the other hand, we note that 
$\phi_c/T_c$ can remain ${\cal O}(1)$ even when
$m_h$ increases, if at the same time $m_\xi$ increases keeping the relation $m_h < m_\xi$. 
In FIG.~\ref{mxi-1F-N8}, a similar plot is shown  for $N_f = 8$. 

In FIG.~\ref{Nf-1F-mxi250}, $\phi_c/T_c$ is plotted as a function of $m_h$ for several values of $N_f$ 
with $m_\xi= 250 \text{GeV}$ fixed. 
For a relatively small  $m_h$, $m_h \lesssim 130 \text{GeV}$, where 
$\phi_c/T_c$ is rather large, we can see a clear dependence on $N_f$: as $N_f$ increases, 
the value of $m_h$ which gives the same value of $\phi_c/T_c$ increases monotonically. 
However, for a relatively large $m_h$,  where $\phi_c/T_c$ gets rather small, the dependence on $N_f$ 
is the opposite. This is because, as one can see from FIG.~\ref{PD-1F-N8}, 
the critical end line shifts towards the stability boundary and  
the region of the first-order transition becomes narrower  as $N_f$ increases. 
As the total result, the value of $m_h$  which gives  $\phi_c/T_c \simeq {\cal O}(1)$ does not
depend on $N_f$ so much.

Throughout the above numerical evaluations of $\phi_c/T_c$,  we observed that 
$T_c/ {\mathcal{M}_i(\phi_c,T_c)} \lesssim {\cal O}(1)$ for $i=h, \pi$, which implies
\begin{eqnarray}
\label{eq:beta-ring-numerical}
\beta_{\lambda_1+\lambda_2/N_f} 
&\simeq& \frac{N_f^2}{4\pi}\left(\lambda_1+\frac{\lambda_2}{N_f}\right), \nonumber\\
\beta_{\lambda_2} 
&\simeq& \frac{N_f^2}{4\pi}\left(\frac{\lambda_2}{N_f}\right) . 
\end{eqnarray}
Then, the ring improved perturbation theory is valid for $\lambda_1+\lambda_2/N_f \lesssim 4\pi/N_f^2$, and, $\lambda_2/N_f \lesssim 4\pi/N_f^2$.  The second condition can be written in the condition of $m_\xi$ as,   
$m_\xi / v \lesssim \sqrt{4\pi/N_f^2}$ for the one-family type 
($m_\xi / v \lesssim \sqrt{2\pi/N_f}$ for the partially gauged type).  
Therefore, one cannot push $m_\xi$ to so large values compared to the weak scale $v$ for a reliable evaluation of $\phi_c/T_c$ within the ring improved perturbation theory. In this paper, we restrict ourselves in the range $m_\xi \lesssim 300  \text{ GeV}$ for both $N_f = 4, 8$.
This may not be safe and the approximation may be rather crude  for $N_f=8$ in particular.
However, this case is also  useful for a comparison and to see the effects of the gauge interactions 
and the symmetry breaking terms.

\subsubsection{SU(2)$\times$U(1) gauged linear sigma model}    
We next show the numerical result for 
the SU(2)$\times$U(1) gauged linear sigma model. 

\begin{figure}[t]
\begin{center}
\includegraphics[width=7.5cm]{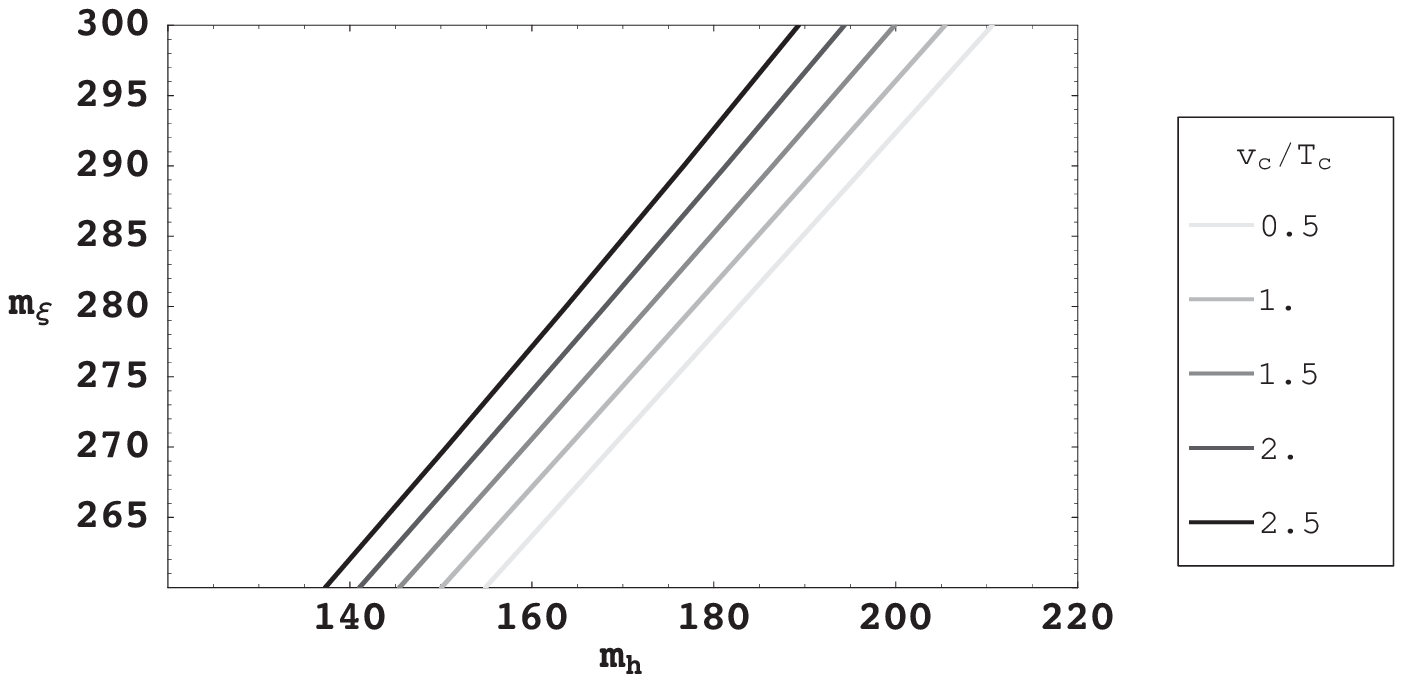}
\caption{Contours of $v_c/T_c$  on the $m_h$-$m_{\xi}$ plane for the one-family model with $N_f=8$.} \label{CP-N8-1F-gs}
\vspace{1em}
\includegraphics[width=7.5cm]{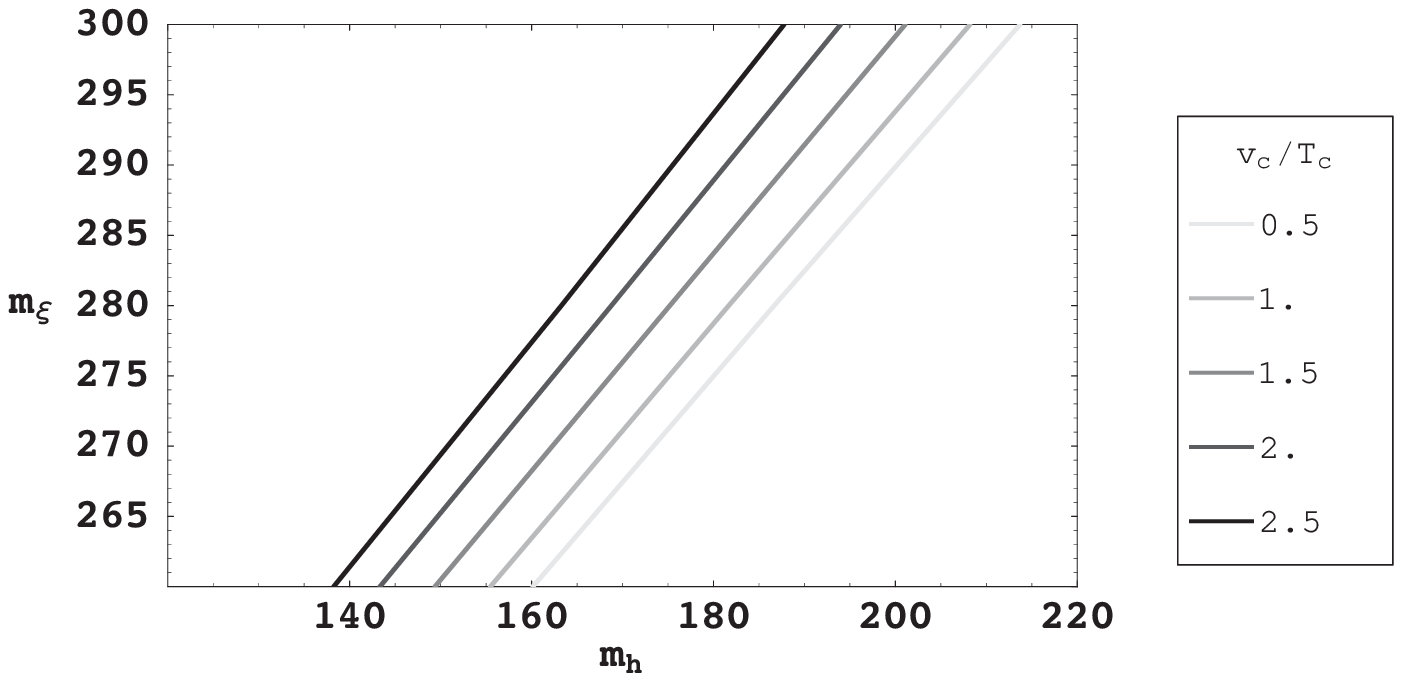}
\caption{Contours of $v_c/T_c$  on the $m_h$-$m_{\xi}$ 
plane for the non-gauged model  with $N_f=8$.
}
\label{CP-N8-1F-s}
\vspace{1em}
\includegraphics[width=7.5cm]{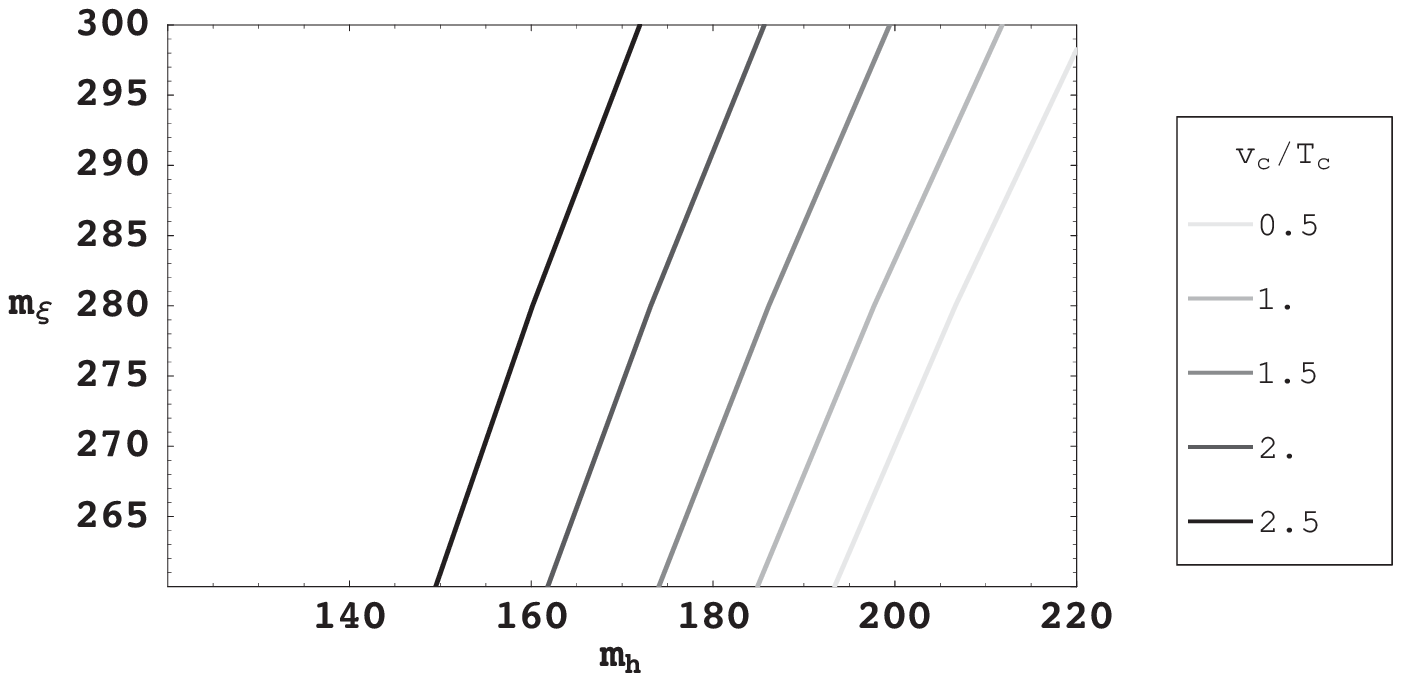}
\caption{Contours of $v_c/T_c$  on the $m_h$-$m_{\xi}$ 
plane for the partially-gauged model with $N_f=8$.
}
\label{CP-N8-1D-gs}
\end{center}
\end{figure}

FIG.~\ref{CP-N8-1F-gs} is a contour plot of $v_c/T_c$ on the $m_h$--$m_\xi$ plane 
for the one-family model with $N_f=8$. 
For a comparison, a similar plot for the non-gauged model with $N_f=8$ 
is shown in FIG.~\ref{CP-N8-1F-s}. 
We can see that the effect of the gauge interactions  slightly reduces
the value of $m_h$ which gives the same value of $v_c/T_c$ for a fixed $m_\xi$.   

FIG.~\ref{CP-N8-1D-gs} is a similar contour plot of $v_c/T_c$ for the partially-gauged
 model with $N_f=8$. 
We note that 
the value of $m_h$ which gives the same value of $v_c/T_c$ for a fixed $m_\xi$ becomes larger in the partially-gauged model.   
This is mainly due to the difference of the relation between $\phi_0$ and  $v$ in this model. 
(See Eq.~(\ref{eq:phi-v-rel-partiall}).)  The partially gauged model is favorable 
than the one-family model in realizing a stronger first-order electroweak phase transition. 


\subsubsection{Explicit symmetry breaking terms} 
Finally, we  show the  numerical result for 
the (non-gauged) linear sigma model with the symmetry breaking terms in the case of $N_f=8$. 
We omit the effect of the gauge interactions because its effect turns out to be relatively small 
compared to the effect of the symmetry breaking terms. 

\begin{figure}[t]
\begin{center}
\includegraphics[width=8cm]{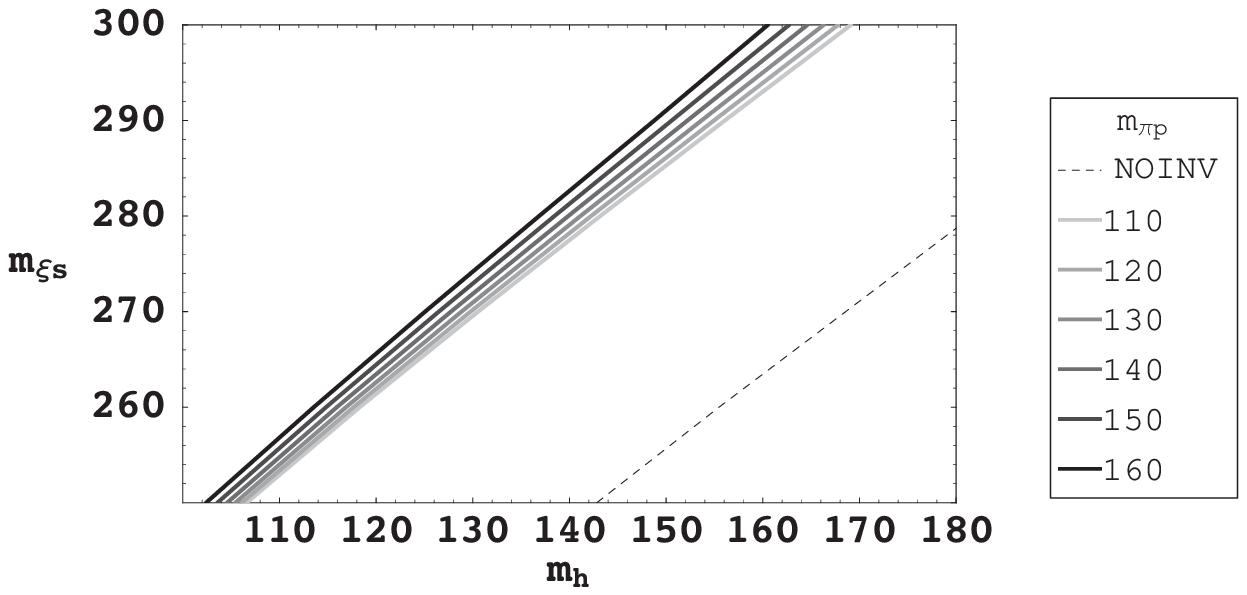}
\caption{
Contours of $\phi_c/T_c = 1$  on the $m_h$-$m_{\xi s}$ plane 
for several values of $m_{\pi p}$ with $m_{\eta} =120 \text{ GeV}$ fixed 
in the non-gauged model with $N_f=8$. 
The dashed line shows the contour in the case without  the breaking term. }
\label{CP-N8-1F-c120}
\vspace{1em}
\includegraphics[width=8cm]{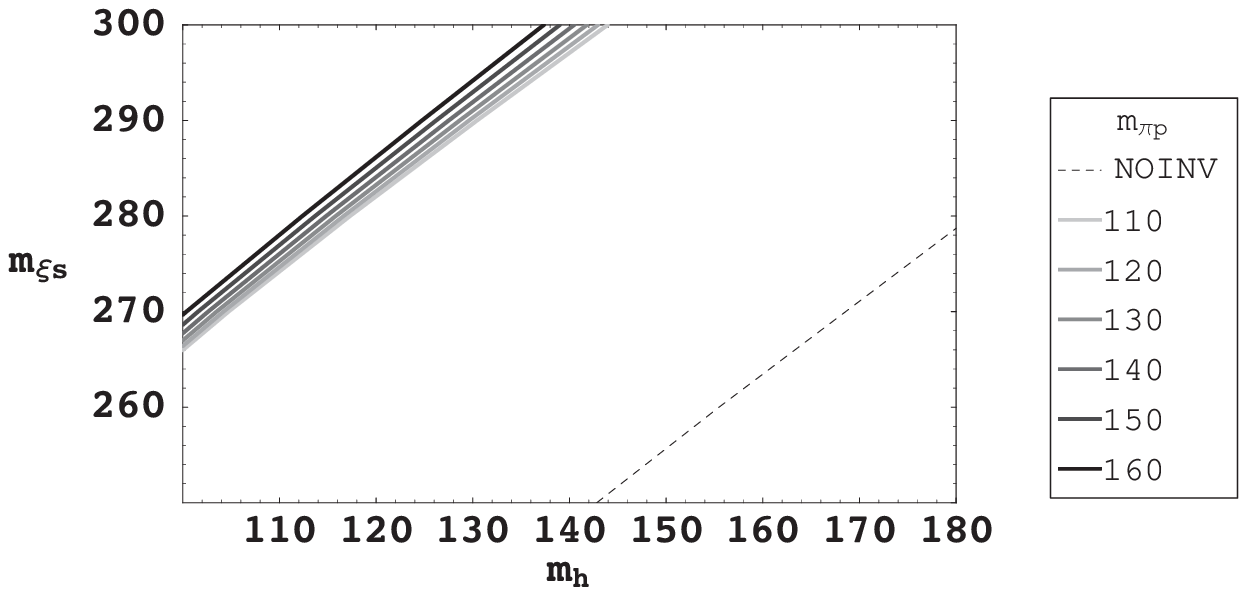}
\caption{
Contours of $\phi_c/T_c = 1$  on the $m_h$-$m_{\xi s}$ plane 
for several values of $m_{\pi p}$ with $m_{\eta} =160 \text{ GeV}$ fixed 
in the non-gauged model with $N_f=8$. 
}
\label{CP-N8-1F-c160}
\end{center}
\end{figure}

%


FIG.~\ref{CP-N8-1F-c120} and FIG.~\ref{CP-N8-1F-c160} show the contours of $\phi_c/T_c = 1$
on the $m_h$--$m_{\xi s}$ plane for several values of the mass parameters, $m_{\pi_P}$, $m_{\eta}$,
of the pseudo NG bosons defined by
\begin{equation}
m_{\eta}^2 = 2c, \qquad m_{\pi_P}^2 = c-\Delta m^2.  
\end{equation} 
We choose the range of the masses as $m_{\eta}, m_{\pi_P} \gtrsim 110 {\rm GeV}$
so that it is within the typically allowed region of  the direct search experiments \cite{Yao:2006px}. (For a specific Extended Technicolor model, the lower bound can be more severe. Taking into account such cases, we searched a larger parameter region. )
We can see that the effect of the symmetry breaking terms
is sizable and reduces the strength of the first-order transition for fixed $m_h$ and $m_{\xi s}$. 
However,  the region of the strongly first-order transition still remains for the allowed values 
of the masses 
of the pseudo NG bosons.  We note also in this case
that $\phi_c/T_c$ can remain ${\cal O}(1)$ even when
$m_h$ increases, if at the same time $m_{\xi s}$ increases keeping the relation $m_h < m_{\xi s}$. 

\section{Summary and Discussion\label{sec:summary-discussion}}
Through the analysis of the finite temperature effective potential at one-loop with the 
ring-improvement, 
we have shown that the electroweak phase transition in the SU(2)$\times$U(1) gauged 
linear sigma model can be strongly first-order with $\phi_c/T_c \simeq {\cal O}(1)$, 
even when the neutral scalar field $h$, as well as the other scalar fields $\xi$,  is rather heavy:
$m_h \simeq 170 {\rm GeV}$, well above the experimental lower bound on the Higgs mass. 
The explicit symmetry breaking terms, which give rise to the masses of the pseudo 
NG  bosons,  reduce the strength of the first-order transition. However, the transition can remain 
strongly first-order 
when the masses of pseudo NG bosons are relatively small and in the range
$m_{\eta}, m_{\pi_P} \gtrsim 110 {\rm GeV}$. 

It is encouraging to observe the fact that 
in order to realize the strongly first-order transition with $\phi_c/T_c \simeq {\cal O}(1)$,
the mass of the neutral scalar field,  $m_h$,  can be rather large
as long as the masses of the extra scalar fields, $m_\xi$, are also large and satisfy the relation
$m_h < m_\xi$. There does not seem to exist the upper bound on $m_h$. 
This does not contradict with 
the possibility that the restoration of the electroweak symmetry is strongly first-order
in the Higgs sector 
possessing the (approximate) SU($N_f$)$\times$SU($N_f$) chiral symmetry with a large $N_f$
as in walking technicolor theories.

It would be quite interesting to explore the critical behavior of the restoration of 
the SU($N_f$)$\times$SU($N_f$) chiral symmetry with a large $N_f$ in massless QCD- like 
theories using numerical  techniques in lattice field theory.  In view of the recent developments of 
the methods for dynamical simulations 
in the chiral regime \cite{Hasenbusch:2001ne, Luscher:2003vf, Luscher:2005rx, Kuramashi:2006np, 
van den Eshof:2002ms,Frommer:1995ik,Chiu:2002eh,Cundy:2004pza,Vranas:2006zk,Fukaya:2006vs,Fukaya:2007fb}, 
this kind of quantitative study from the first principle should become available in quite near future.  
See \cite{Maezawa:2007fd,Ukita:2006pc} for the current status of the numerical study 
of two-flavor QCD. 

It would be also interesting to examine the SU(4)/O(4) chiral symmetry restoration at finite temperature
from the point of view of the electroweak phase transition.  This pattern of the  chiral symmetry breaking / restoration appears in the minimal walking techicolor theory \cite{Sannino:2004qp,Dietrich:2006cm}. 
In the linear sigma model of this case, the determiant term which breaks the U(1) chiral symmetry is relevant and gives rise to another quartic coupling.  
It has been argued in \cite{Basile:2004wa} that when the derminant term is large, 
there is a stable IR fixed point which corresponds to a new three-dimensional universality class characterized by the symmetry breaking pattern SU(4)/O(4).  (This fixed point does not appear in the $\epsilon$-expansion.) This implies that the finite temperature restoration of the chiral symmetry may be continuous, 
although it does not exclude a first-order transition for the system that are outside the
domain of the stable fixed point. 
In order to get some insights on the interplay between 
the low-lying spectrum  in the minimal walking techicolor theory 
and the nature of the critical behavior of the chiral symmetry restoration, 
a semi-quantitative analysis within the linear sigma model using the finite temperature 
effective potential may be useful. 
Results of the numerical studies using lattice field theory has been reported 
in \cite{Karsch:1998qj,Engels:2005te}.
See also \cite{Catterall:2007yx} for a recent attempt to observe the "walking" behavior in the model. 

In our analysis of the SU(2)$\times$U(1) gauged linear sigma model, we have used 
the sphaleron solution to estimate the rate of the baryon number violating interactions. 
In the case of massless QCD-like theories coupled to the SU(2)$\times$U(1) electroweak interactions, 
it seems non-trivial  to estimate the rate by identifying the counterpart of the sphaleron solution. 
We leave this question for a future study.

\vspace{2em}
\begin{acknowledgments}
The authors would like to thank K.~Funakubo, K.~Yamawaki, M.~Harada, N.~Maekawa, K.~Sakurai 
for valuable discussions.
Y.K. is supported in part by Grant-in-Aid for Scientific Research No.~17540249.
\end{acknowledgments}

\appendix

\section{Effective T-dependent Mass in SU(2)$\times$U(1) linear sigma model} 

In this appendix, we show the effective T-dependent masses (mass matrices) for the scalar bosons and 
the longitudinal components of the gauge bosons in the gauged linear sigma models of the one-family type with $N_f=8$ and patialy-gauged type with $N_f\geq 3$.
In the calculation of the scalar boson self energies and the gauge boson vacuum polarizatoins, we take the IR limit where the Matsubara frequency and 
the momentum of the external fields go to zero and retain only the leading order terms of $m_i(\phi)/T$.
We also show the corresponding degrees of freedom $n_i$.

\subsection{partialy-gauged type}
As the scalar field $\Phi$ partialy couples to the gauge fields in this case, it is useful to decompose the field $\Phi$ as follows:   
\begin{align}
\Phi =&\left( 
\begin{array}{@{\,}c|c@{\,}} 
\frac{\tilde h}{2}\openone_2+\tilde\Xi 
& \frac{1}{\sqrt{2}}\begin{pmatrix} \xi_{13} & \dots & \xi_{1N_f} \\ \xi_{23} & \dots & \xi_{2N_f} \end{pmatrix} \\ \hline
\frac{1}{\sqrt{2}}\begin{pmatrix} \xi_{13}^* & \xi_{23}^* \\ \vdots & \vdots \\ \xi_{1N_f}^* & \xi_{2N_f}^* \end{pmatrix} 
& \frac{\check h}{\sqrt{2(N_f-2)}}\openone_{N_f-2}+\check\Xi
\end{array}
\right)  \notag \\
&+i\left( 
\begin{array}{@{\,}c|c@{\,}} 
\frac{\tilde \eta}{2}\openone_2+\tilde\Pi 
& \frac{1}{\sqrt{2}}\begin{pmatrix} \pi_{13} & \dots & \pi_{1N_f} \\ \pi_{23} & \dots & \pi{2N_f} \end{pmatrix} \\ \hline
\frac{1}{\sqrt{2}}\begin{pmatrix} \pi_{13}^* & \pi_{23}^* \\ \vdots & \vdots \\ \pi_{1N_f}^* & \pi_{2N_f}^* \end{pmatrix} 
& \frac{\check \eta}{\sqrt{2(N_f-2)}}\openone_{N_f-2}+\check\Pi
\end{array}
\right), 
\end{align}
where $\tilde h,~\check h,~\tilde \eta$ and $\check \eta$ are real fields, $\tilde \Xi$ and $\tilde \Pi$ are $2\times2$ traceless Hermitian matrix valued fields, 
$\check \Xi$ and $\check \Pi$ are $(N_f-2)\times(N_f-2)$ traceless Hermitian matrix valued fields, 
$\xi_{\alpha\bar\beta}$ and $\pi_{\alpha\bar\beta}~(\alpha=1,2;\bar\beta=3,\dots ,N_f )$ are complex fields, 
and $\openone_{N_f-2}$ is $(N_f-2)\times(N_f-2)$ unit matrix. 

The one-loop contributions to the scalar boson self energies are shown in FIG.~\ref{fig:1D-loop}.
We use the short-hand notation
\begin{align}
\Pi^{(S)}&=\frac{T^2}{12}[(N_f^2+1)\lambda_1+2N_f\lambda_2],\notag\\
\Pi^{(2)}&=\frac{3}{16}g^2T^2~,~\Pi^{(1)}=\frac{1}{16}g^{\prime 2}T^2, \label{eq:self_sab}
\end{align}
which are corresponding to the contributions to the scalar boson self energies from the scalar bosons themselves and SU(2), U(1) gauge bosons, respectively. 

The effective T-dependent mass matrix for $\tilde h$ and $\check h$ is
\begin{widetext}
\begin{equation}
\begin{pmatrix}  
\frac{2}{N_f}m_h^2(\phi)+\frac{N_f-2}{N_f}m_\xi^2(\phi)+\Pi^{(S)}+\Pi^{(1)}+\Pi^{(2)} 
& \frac{\sqrt{2(N_f-2)}}{N_f}(m_h^2(\phi)-m_\xi^2(\phi)) \\
\frac{\sqrt{2(N_f-2)}}{N_f}(m_h^2(\phi)-m_\xi^2(\phi)) &
\frac{N_f-2}{N_f}m_h^2(\phi)+\frac{2}{N_f}m_\xi^2(\phi)+\Pi^{(S)}
\end{pmatrix}, 
\end{equation}
and each mass eigenvalue has one degree of freedom. 
The effective T-dependent mass matrix for $\xi_{\alpha\bar\beta}$ and $\pi_{\alpha\bar\beta}$ is
\begin{equation}
\begin{pmatrix}  
m_\xi^2(\phi)+\Pi^{(S)}+\frac{1}{2}(\Pi^{(1)}+\Pi^{(2)}) &
-\frac{i}{2}(\Pi^{(1)}-\Pi^{(2)}) \\
\frac{i}{2}(\Pi^{(1)}-\Pi^{(2)}) &
m_\pi^2(\phi)+\Pi^{(S)}+\frac{1}{2}(\Pi^{(1)}+\Pi^{(2)}) 
\end{pmatrix}, 
\end{equation}
\end{widetext}
and each mass eigenvalue has $4(N_f-2)$ degrees of freedom.
The other effective T-dependent masses for the scalar bosons are given by
\begin{gather}
\mathcal{M}_{\tilde h}^2(\phi,T)=\mathcal{M}_{\tilde\xi}^2(\phi,T)=m_\xi^2(\phi)+\Pi^{(S)}+\Pi^{(1)}+\Pi^{(2)}, \notag \\
\mathcal{M}_{\check \xi}^2(\phi,T)=m_\xi^2(\phi)+\Pi^{(S)}, \notag \\
\mathcal{M}_{\tilde \eta}^2(\phi,T)=\mathcal{M}_{\tilde\pi}^2(\phi,T)=m_\pi^2(\phi)+\Pi^{(S)}+\Pi^{(1)}+\Pi^{(2)}, \notag\\
\mathcal{M}_{\check \pi}^2(\phi,T)=m_\pi^2(\phi)+\Pi^{(S)}, 
\end{gather}
and the corresponding degrees of freedom are
\begin{gather}
n_{\tilde h}=n_{\tilde \eta}=1,~n_{\tilde\xi}=n_{\tilde \pi}=3, \notag \\
n_{\check \xi}=n_{\check \pi}=(N_f-2)^2-1.
\end{gather}

For the longitudinal components of the gauge bosons, the one-loop contributions to the vacuum polarizations are shown in FIG.~\ref{fig:gloop}.
For the W boson,
\begin{gather}
\mathcal{M}_{W_L}^2(\phi ,T)=\frac{1}{2N_f}g^2\phi^2+\frac{N_f+4}{6}g^2T^2,\notag\\
n_{W_L}=2.
\end{gather}
The effective T-dependent mass matrix for the Z boson and the photon, in the ($A^3,B$) basis, is 
\begin{align}
\begin{pmatrix} 
\frac{1}{2N_f}g^2\phi^2+\frac{N_f+4}{6}g^2T^2 & -\frac{1}{2N_f}gg^{\prime}\phi^2 \\
-\frac{1}{2N_f}gg^{\prime}\phi^2 & \frac{1}{2N_f}g^{\prime 2}\phi^2+\frac{N_f}{6}g^{\prime2}T^2
\end{pmatrix}, 
\end{align}
and each mass eigenvalue has one degree of freedom.

\subsection{one-family type with $N_f=8$}
\begin{figure}
  \centering
  \includegraphics[width=200pt,clip]{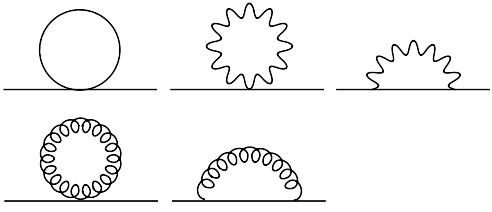}
  \caption{The one-loop self energy diagrams of scalar bosons in the one family model.}
  \label{fig:1F-loop}
\end{figure}

The one-loop contributions to the scalar boson self energies are shown in FIG.~\ref{fig:1F-loop}.
We use the notation (\ref{eq:self_sab}) again, for the scalar boson loop, W boson loop and Z boson loop contributions to the scalar boson self-energies.%
In this model, there is the additional contribution from the gluon loop and we also use the short-hand notation
\begin{gather}
\Pi^{(3)}(\bold{8})=2g_s^2T^2,~\Pi^{(3)}(\bold{3})=\frac{1}{3}g_s^2T^2, 
\end{gather}
which are corresponding to the contributions to the color octet and the color triplet scalar boson self energies from the gluon, respectively. 

\begin{widetext}
We show the effective T-dependent masses (and mass matrices) for the scalar bosons on the color irreducible basis.
\begin{itemize}
\item Color singlet $S$ and $S^\prime$:
\begin{gather}
\mathcal{M}_{h_S}^2(\phi ,T)=m_h^2(\phi)+\Pi^{(S)}+\Pi^{(2)}+\Pi^{(1)};~n_{h_S}=1, \notag\\
\mathcal{M}_{h_{S^\prime}}^2(\phi ,T)=\mathcal{M}_{\xi_S}^2(\phi ,T)=\mathcal{M}_{\xi_{S^\prime}}^2(\phi ,T)=m_\xi^2(\phi)+\Pi^{(S)}+\Pi^{(2)}+\Pi^{(1)};~ 
n_{h_{S^\prime}}=1,~n_{\xi_S}=n_{\xi_{S^\prime}}=3, \notag \\
\mathcal{M}_{\eta_S}^2(\phi ,T)=\mathcal{M}_{\eta_{S^\prime}}^2(\phi ,T)=m_\eta^2(\phi)+\Pi^{(S)}+\Pi^{(2)}+\Pi^{(1)};~ n_{\eta_S}=n_{\eta_{S^\prime}}=1, \notag \\
\mathcal{M}_{\pi_S}^2(\phi ,T)=\mathcal{M}_{\pi_{S^\prime}}^2(\phi ,T)=m_\pi^2(\phi)+\Pi^{(S)}+\Pi^{(2)}+\Pi^{(1)};~ 
n_{\pi_S}=n_{\pi_{S^\prime}}=3.
\end{gather}

\item Color octet $O$:
\begin{gather}
\mathcal{M}_{h_O}^2(\phi ,T)=\mathcal{M}_{\xi_O}^2(\phi ,T)=m_\xi^2(\phi)+\Pi^{(S)}+\Pi^{(3)}(\bold{8})+\Pi^{(2)}+\Pi^{(1)};~ 
n_{h_O}=8,~n_{\xi_O}=24, \notag \\
\mathcal{M}_{\eta_O}^2(\phi ,T)=\mathcal{M}_{\pi_O}^2(\phi ,T)=m_\pi^2(\phi)+\Pi^{(S)}+\Pi^{(3)}(\bold{8})+\Pi^{(2)}+\Pi^{(1)};~  
n_{\eta_O}=8,~n_{\pi_O}=24.
\end{gather}

\item Color triplet $T$ and anti-triplet $\bar{T}$:

The mass term induce the mixing between $T$ and $\bar{T}$.

There are four effective T-dependent mass matrices: 
\begin{equation}
\begin{pmatrix}
\frac{m_\xi^2(\phi)+m_\pi^2(\phi)}{2} + \Pi(\bold{3}) & -\frac{24}{9}\Pi^{(1)} & \frac{m_\xi^2(\phi)-m_\pi^2(\phi)}{2} & 0 \\
-\frac{24}{9}\Pi^{(1)} & \frac{m_\xi^2(\phi)+m_\pi^2(\phi)}{2} + \Pi(\bold{3}) & 0 & \frac{m_\xi^2(\phi)-m_\pi^2(\phi)}{2} \\
\frac{m_\xi^2(\phi)-m_\pi^2(\phi)}{2} & 0 & \frac{m_\xi^2(\phi)+m_\pi^2(\phi)}{2} + \Pi(\bold{3}) & -\frac{24}{9}\Pi^{(1)} \\
0 & \frac{m_\xi^2(\phi)-m_\pi^2(\phi)}{2} & -\frac{24}{9}\Pi^{(1)} & \frac{m_\xi^2(\phi)+m_\pi^2(\phi)}{2} + \Pi(\bold{3}) 
\end{pmatrix}
\end{equation}
for $(h_T^i,\xi_T^{i,3},h_{\bar{T}}^i,\xi_{\bar{T}}^{i,3})$, and

\begin{equation}
\begin{pmatrix}
\frac{m_\xi^2(\phi)+m_\pi^2(\phi)}{2} + \Pi(\bold{3}) & -\frac{24}{9}\Pi^{(1)} & -\frac{m_\xi^2(\phi)-m_\pi^2(\phi)}{2} & 0 \\
-\frac{24}{9}\Pi^{(1)} & \frac{m_\xi^2(\phi)+m_\pi^2(\phi)}{2} + \Pi(\bold{3}) & 0 & -\frac{m_\xi^2(\phi)-m_\pi^2(\phi)}{2} \\
-\frac{m_\xi^2(\phi)-m_\pi^2(\phi)}{2} & 0 & \frac{m_\xi^2(\phi)+m_\pi^2(\phi)}{2} + \Pi(\bold{3}) & -\frac{24}{9}\Pi^{(1)} \\
0 & -\frac{m_\xi^2(\phi)-m_\pi^2(\phi)}{2} & -\frac{24}{9}\Pi^{(1)} & \frac{m_\xi^2(\phi)+m_\pi^2(\phi)}{2} + \Pi(\bold{3}) 
\end{pmatrix}
\end{equation}
for $(\eta_T^i,\pi_T^{i,3},\eta_{\bar{T}}^i,\pi_{\bar{T}}^{i,3})$, and 

\begin{equation}
\begin{pmatrix}
\frac{m_\xi^2(\phi)+m_\pi^2(\phi)}{2} + \Pi(\bold{3}) & \frac{24}{9}\Pi^{(1)} & \frac{m_\xi^2(\phi)-m_\pi^2(\phi)}{2} & 0 \\
\frac{24}{9}\Pi^{(1)} & \frac{m_\xi^2(\phi)+m_\pi^2(\phi)}{2} + \Pi(\bold{3}) & 0 & -\frac{m_\xi^2(\phi)-m_\pi^2(\phi)}{2} \\
\frac{m_\xi^2(\phi)-m_\pi^2(\phi)}{2} & 0 & \frac{m_\xi^2(\phi)+m_\pi^2(\phi)}{2} + \Pi(\bold{3}) & -\frac{24}{9}\Pi^{(1)} \\
0 & -\frac{m_\xi^2(\phi)-m_\pi^2(\phi)}{2} & -\frac{24}{9}\Pi^{(1)} & \frac{m_\xi^2(\phi)+m_\pi^2(\phi)}{2} + \Pi(\bold{3}) 
\end{pmatrix}
\end{equation}
for $(\xi_T^{i,1},\pi_T^{i,2},\xi_{\bar{T}}^{i,1},\pi_{\bar{T}}^{i,2})$, and

\begin{equation}
\begin{pmatrix}
\frac{m_\xi^2(\phi)+m_\pi^2(\phi)}{2} + \Pi(\bold{3}) & -\frac{24}{9}\Pi^{(1)} & \frac{m_\xi^2(\phi)-m_\pi^2(\phi)}{2} & 0 \\
-\frac{24}{9}\Pi^{(1)} & \frac{m_\xi^2(\phi)+m_\pi^2(\phi)}{2} + \Pi(\bold{3}) & 0 & -\frac{m_\xi^2(\phi)-m_\pi^2(\phi)}{2} \\
\frac{m_\xi^2(\phi)-m_\pi^2(\phi)}{2} & 0 & \frac{m_\xi^2(\phi)+m_\pi^2(\phi)}{2} + \Pi(\bold{3}) & -\frac{24}{9}\Pi^{(1)} \\
0 & -\frac{m_\xi^2(\phi)-m_\pi^2(\phi)}{2} & \frac{24}{9}\Pi^{(1)} & \frac{m_\xi^2(\phi)+m_\pi^2(\phi)}{2} + \Pi(\bold{3}) 
\end{pmatrix}
\end{equation}
for $(\xi_T^{i,2},\pi_T^{i,1},\xi_{\bar{T}}^{i,2},\pi_{\bar{T}}^{i,1})$, where 
\begin{equation}
\Pi(\bold{3})= \Pi^{S}+\Pi^{(3)}(\bold{3})+\Pi^{(2)}+\frac{25}{9}\Pi^{(1)}.
\end{equation}
These four matrices have same eigenvalues. 
After the diagonalization of these mass matrices, we obtain the effective T-dependent masses for the color (anti-) triplet scalar bosons.
The eigenvalues are
\begin{equation}
m_{\xi}^2(\phi)+\Pi(\bold{3})\pm\frac{8}{3}\Pi^{(1)},~m_{\pi}^2(\phi)+\Pi(\bold{3})\pm\frac{8}{3}\Pi^{(1)}, 
\end{equation}
and each mass eigenvalue has  12 degrees of freedom.

\end{itemize}
\end{widetext}

For the longitudinal components of the gauge bosons, the one-loop contributions to the vacuum polarizations are shown in FIG.~\ref{fig:gloop}.
For the W boson,
\begin{equation}
\mathcal{M}_{W_L}^2(\phi ,T)=\frac{1}{4}g^2\phi^2+6g^2T^2,~n_{W_L}=2.
\end{equation}
The effective T-dependent mass matrix for the Z boson and the photon, in the ($A^3,B$) basis, is 
\begin{align}
\begin{pmatrix} 
\frac{1}{4}g^2\phi^2+6g^2T^2 & -\frac{1}{4}gg^{\prime}\phi^2 \\
-\frac{1}{4}gg^{\prime}\phi^2 & \frac{1}{4}g^{\prime 2}\phi^2+\frac{80}{9}g^{\prime2}T^2
\end{pmatrix}, 
\end{align}
and each mass eigenvalue has one degree of freedom. \\



\begin{thebibliography}{99}

\bibitem{Sakharov:1973}
A.D.~Sakharov, Pis'ma Zh. Eksp. Teor. Fiz. {\bf 5}, 32 (1967)  [JETP Lett. {\bf 5}, 24 (1967)]

\bibitem{Kuzmin:1985mm}
  V.~A.~Kuzmin, V.~A.~Rubakov and M.~E.~Shaposhnikov,
  Phys.\ Lett.\  B {\bf 155}, 36 (1985).
  
\bibitem{Cohen:1990py}
  A.~G.~Cohen, D.~B.~Kaplan and A.~E.~Nelson,
  Phys.\ Lett.\  B {\bf 245}, 561 (1990).
  
\bibitem{Cohen:1990it}
  A.~G.~Cohen, D.~B.~Kaplan and A.~E.~Nelson,
  Nucl.\ Phys.\  B {\bf 349}, 727 (1991).

\bibitem{Spergel:2006hy}
  D.~N.~Spergel {\it et al.}  [WMAP Collaboration],
  Astrophys.\ J.\ Suppl.\  {\bf 170}, 377 (2007)
  [arXiv:astro-ph/0603449].

\bibitem{Steigman:2005uz}
  G.~Steigman,
  Int.\ J.\ Mod.\ Phys.\  E {\bf 15}, 1 (2006)
  [arXiv:astro-ph/0511534].
  
  

\bibitem{Kobayashi:1973fv}
  M.~Kobayashi and T.~Maskawa,
  Prog.\ Theor.\ Phys.\  {\bf 49}, 652 (1973).

%


\bibitem{Shaposhnikov:1987tw}
  M.~E.~Shaposhnikov,
  Nucl.\ Phys.\ B {\bf 287}, 757 (1987).

\bibitem{Farrar:1993sp}
  G.~R.~Farrar and M.~E.~Shaposhnikov,
  Phys.\ Rev.\ Lett.\  {\bf 70}, 2833 (1993)
  [Erratum-ibid.\  {\bf 71}, 210 (1993)]
  [arXiv:hep-ph/9305274].

\bibitem{Farrar:1993hn}
  G.~R.~Farrar and M.~E.~Shaposhnikov,
  Phys.\ Rev.\ D {\bf 50}, 774 (1994)
  [arXiv:hep-ph/9305275].

\bibitem{Gavela:1994dt}
  M.~B.~Gavela, P.~Hernandez, J.~Orloff, O.~Pene and C.~Quimbay,
  Nucl.\ Phys.\ B {\bf 430}, 382 (1994)
  [arXiv:hep-ph/9406289].

\bibitem{Gavela:1993ts}
  M.~B.~Gavela, P.~Hernandez, J.~Orloff and O.~Pene,
  Mod.\ Phys.\ Lett.\ A {\bf 9}, 795 (1994)
  [arXiv:hep-ph/9312215].

\bibitem{Huet:1994jb}
  P.~Huet and E.~Sather,
  Phys.\ Rev.\ D {\bf 51}, 379 (1995)
  [arXiv:hep-ph/9404302].


\bibitem{:2003ih}
    [LEP Collaboration],
  arXiv:hep-ex/0312023.

\bibitem{Kajantie:1996mn}
  K.~Kajantie, M.~Laine, K.~Rummukainen and M.~E.~Shaposhnikov,
  Phys.\ Rev.\ Lett.\  {\bf 77}, 2887 (1996)
  [arXiv:hep-ph/9605288].

\bibitem{Rummukainen:1998as}
  K.~Rummukainen, M.~Tsypin, K.~Kajantie, M.~Laine and M.~E.~Shaposhnikov,
  Nucl.\ Phys.\ B {\bf 532}, 283 (1998)
  [arXiv:hep-lat/9805013].

\bibitem{Csikor:1998eu}
  F.~Csikor, Z.~Fodor and J.~Heitger,
  Phys.\ Rev.\ Lett.\  {\bf 82}, 21 (1999)
  [arXiv:hep-ph/9809291].
  
\bibitem{Aoki:1999fi}
  Y.~Aoki, F.~Csikor, Z.~Fodor and A.~Ukawa,
  Phys.\ Rev.\  D {\bf 60}, 013001 (1999)
  [arXiv:hep-lat/9901021].
  
  

\bibitem{Bochkarev:1990fx}
  A.~I.~Bochkarev, S.~V.~Kuzmin and M.~E.~Shaposhnikov,
  Phys.\ Lett.\  B {\bf 244}, 275 (1990).

\bibitem{Bochkarev:1990gb}
  A.~I.~Bochkarev, S.~V.~Kuzmin and M.~E.~Shaposhnikov,
  Phys.\ Rev.\  D {\bf 43}, 369 (1991).

\bibitem{Cohen:1991iu}
  A.~G.~Cohen, D.~B.~Kaplan and A.~E.~Nelson,
  Phys.\ Lett.\  B {\bf 263}, 86 (1991).

\bibitem{Nelson:1991ab}
  A.~E.~Nelson, D.~B.~Kaplan and A.~G.~Cohen,
  Nucl.\ Phys.\  B {\bf 373}, 453 (1992).


\bibitem{Turok:1990zg}
  N.~Turok and J.~Zadrozny,
  Nucl.\ Phys.\  B {\bf 358}, 471 (1991).
%

\bibitem{Turok:1991uc}
  N.~Turok and J.~Zadrozny,
  Nucl.\ Phys.\  B {\bf 369}, 729 (1992).
  
  
\bibitem{Funakubo:1993jg}
  K.~Funakubo, A.~Kakuto and K.~Takenaga,
  Prog.\ Theor.\ Phys.\  {\bf 91}, 341 (1994)
  [arXiv:hep-ph/9310267].

\bibitem{Davies:1994id}
  A.~T.~Davies, C.~D.~Froggatt, G.~Jenkins and R.~G.~Moorhouse,
  Phys.\ Lett.\  B {\bf 336}, 464 (1994).

\bibitem{Cline:1995dg}
  J.~M.~Cline, K.~Kainulainen and A.~P.~Vischer,
  Phys.\ Rev.\  D {\bf 54}, 2451 (1996)
  [arXiv:hep-ph/9506284].
  
\bibitem{Cline:1996mg}
  J.~M.~Cline and P.~A.~Lemieux,
  Phys.\ Rev.\ D {\bf 55}, 3873 (1997)
  [arXiv:hep-ph/9609240].
  
\bibitem{Fromme:2006cm}
  L.~Fromme, S.~J.~Huber and M.~Seniuch,
  JHEP {\bf 0611}, 038 (2006)
  [arXiv:hep-ph/0605242].
  

\bibitem{Kanemura:2004ch}
  S.~Kanemura, Y.~Okada and E.~Senaha,
  Phys.\ Lett.\  B {\bf 606}, 361 (2005)
  [arXiv:hep-ph/0411354].


\bibitem{Carena:1996wj}
  M.~S.~Carena, M.~Quiros and C.~E.~M.~Wagner,
  Phys.\ Lett.\  B {\bf 380}, 81 (1996)
  [arXiv:hep-ph/9603420].
  
\bibitem{Cline:1996cr}
  J.~M.~Cline and K.~Kainulainen,
  Nucl.\ Phys.\  B {\bf 482}, 73 (1996)
  [arXiv:hep-ph/9605235].
%
%

\bibitem{Laine:1996ms}
  M.~Laine,
  Nucl.\ Phys.\  B {\bf 481}, 43 (1996)
  [Erratum-ibid.\  B {\bf 548}, 637 (1999)]
  [arXiv:hep-ph/9605283].
  
\bibitem{Losada:1996ju}
  M.~Losada,
  Phys.\ Rev.\  D {\bf 56} (1997) 2893
  [arXiv:hep-ph/9605266].
  
%
%
  
\bibitem{Cline:1998hy}
  J.~M.~Cline and G.~D.~Moore,
  Phys.\ Rev.\ Lett.\  {\bf 81}, 3315 (1998)
  [arXiv:hep-ph/9806354].

\bibitem{Laine:1998vn}
  M.~Laine and K.~Rummukainen,
  Phys.\ Rev.\ Lett.\  {\bf 80}, 5259 (1998)
  [arXiv:hep-ph/9804255].
  
\bibitem{Laine:1998qk}
  M.~Laine and K.~Rummukainen,
  Nucl.\ Phys.\  B {\bf 535}, 423 (1998)
  [arXiv:hep-lat/9804019].
 
%
\bibitem{Cline:1997vk}
  J.~M.~Cline, M.~Joyce and K.~Kainulainen,
  Phys.\ Lett.\  B {\bf 417}, 79 (1998)
  [Erratum-ibid.\  B {\bf 448}, 321 (1999)]
  [arXiv:hep-ph/9708393].

\bibitem{Huet:1995sh}
  P.~Huet and A.~E.~Nelson,
  Phys.\ Rev.\  D {\bf 53}, 4578 (1996)
  [arXiv:hep-ph/9506477].


\bibitem{Carena:1997gx}
  M.~S.~Carena, M.~Quiros, A.~Riotto, I.~Vilja and C.~E.~M.~Wagner,
  Nucl.\ Phys.\  B {\bf 503}, 387 (1997)
  [arXiv:hep-ph/9702409].
  
\bibitem{Joyce:1999fw}
  M.~Joyce, K.~Kainulainen and T.~Prokopec,
  Phys.\ Lett.\  B {\bf 468}, 128 (1999)
  [arXiv:hep-ph/9906411].
  
\bibitem{Joyce:2000ed}
  M.~Joyce, K.~Kainulainen and T.~Prokopec,
  JHEP {\bf 0010}, 029 (2000)
  [arXiv:hep-ph/0002239].

\bibitem{Kainulainen:2001cn}
  K.~Kainulainen, T.~Prokopec, M.~G.~Schmidt and S.~Weinstock,
  JHEP {\bf 0106}, 031 (2001)
  [arXiv:hep-ph/0105295].
   
\bibitem{Kainulainen:2002th}
  K.~Kainulainen, T.~Prokopec, M.~G.~Schmidt and S.~Weinstock,
  Phys.\ Rev.\  D {\bf 66}, 043502 (2002)
  [arXiv:hep-ph/0202177].

 
\bibitem{Funakubo:1998jz}
  K.~Funakubo, A.~Kakuto, S.~Otsuki and F.~Toyoda,
  Prog.\ Theor.\ Phys.\  {\bf 99}, 1045 (1998)
  [arXiv:hep-ph/9802276].
 
\bibitem{Funakubo:1998fk}
  K.~Funakubo,
  Prog.\ Theor.\ Phys.\  {\bf 101}, 415 (1999)
  [arXiv:hep-ph/9809517].

\bibitem{Funakubo:1999ws}
  K.~Funakubo, S.~Otsuki and F.~Toyoda,
  Prog.\ Theor.\ Phys.\  {\bf 102}, 389 (1999)
  [arXiv:hep-ph/9903276].
 %
 
\bibitem{Funakubo:2002yb}
  K.~Funakubo, S.~Tao and F.~Toyoda,
  Prog.\ Theor.\ Phys.\  {\bf 109}, 415 (2003)
  [arXiv:hep-ph/0211238].

  
%
%


\bibitem{Pietroni:1992in}
  M.~Pietroni,
  Nucl.\ Phys.\  B {\bf 402}, 27 (1993)
  [arXiv:hep-ph/9207227].
  
\bibitem{Davies:1996qn}
  A.~T.~Davies, C.~D.~Froggatt and R.~G.~Moorhouse,
  Phys.\ Lett.\  B {\bf 372}, 88 (1996)
  [arXiv:hep-ph/9603388].
 
\bibitem{Huber:1998ck}
  S.~J.~Huber and M.~G.~Schmidt,
  Eur.\ Phys.\ J.\  C {\bf 10}, 473 (1999)
  [arXiv:hep-ph/9809506].
  
\bibitem{Huber:2000mg}
  S.~J.~Huber and M.~G.~Schmidt,
  Nucl.\ Phys.\  B {\bf 606}, 183 (2001)
  [arXiv:hep-ph/0003122].


\bibitem{Kang:2004pp}
  J.~Kang, P.~Langacker, T.~j.~Li and T.~Liu,
  Phys.\ Rev.\ Lett.\  {\bf 94}, 061801 (2005)
  [arXiv:hep-ph/0402086].

\bibitem{Funakubo:2004ka}
  K.~Funakubo and S.~Tao,
  Prog.\ Theor.\ Phys.\  {\bf 113}, 821 (2005)
  [arXiv:hep-ph/0409294].
  
\bibitem{Funakubo:2005pu}
  K.~Funakubo, S.~Tao and F.~Toyoda,
  Prog.\ Theor.\ Phys.\  {\bf 114}, 369 (2005)
  [arXiv:hep-ph/0501052].
  
\bibitem{Funakubo:2005bu}
  K.~Funakubo, A.~Kakuto, S.~Tao and F.~Toyoda,
  Prog.\ Theor.\ Phys.\  {\bf 114}, 1069 (2006)
  [arXiv:hep-ph/0506156].
  
\bibitem{Huber:2006wf}
  S.~J.~Huber, T.~Konstandin, T.~Prokopec and M.~G.~Schmidt,
  Nucl.\ Phys.\  B {\bf 757}, 172 (2006)
  [arXiv:hep-ph/0606298].
  
\bibitem{Grojean:2004xa}
  C.~Grojean, G.~Servant and J.~D.~Wells,
  Phys.\ Rev.\ D {\bf 71}, 036001 (2005)
  [arXiv:hep-ph/0407019].
  
\bibitem{Ham:2004zs}
  S.~W.~Ham and S.~K.~Oh,
  Phys.\ Rev.\ D {\bf 70}, 093007 (2004)
  [arXiv:hep-ph/0408324].
  
\bibitem{Bodeker:2004ws}
  D.~Bodeker, L.~Fromme, S.~J.~Huber and M.~Seniuch,
  JHEP {\bf 0502}, 026 (2005)
  [arXiv:hep-ph/0412366].

  
\bibitem{Coleman:1973jx}
  S.~R.~Coleman and E.~Weinberg,
  Phys.\ Rev.\  D {\bf 7}, 1888 (1973).

\bibitem{Bak-Krinsky-Mukamel:1976}
P.~Bak, S.~Krinsky and D.~Mukamel, Phys.\ Rev.\ Lett.\  {\bf 36}, 52 (1976)
 
\bibitem{Iacobson:1981jm}
  H.~H.~Iacobson and D.~J.~Amit,
  Annals Phys.\  {\bf 133}, 57 (1981).

\bibitem{Pisarski:1983ms}
  R.~D.~Pisarski and F.~Wilczek,
  Phys.\ Rev.\ D {\bf 29}, 338 (1984).
     
\bibitem{Rudnick:1978}
J.~Rudnic, Phys.\ Rev.\ B {\bf 18}, 1406 (1978). 



\bibitem{Amit:1978dk}
  D.~J.~Amit,
``Field Theory, The Renormalization Group, And Critical Phenomena,''
revised second edition, {\it  World Scientific  1984, 394p}

\bibitem{Ginsparg:1980ef}
  P.~H.~Ginsparg,
  Nucl.\ Phys.\  B {\bf 170}, 388 (1980).

\bibitem{Yamagishi:1981qq}
  H.~Yamagishi,
  Phys.\ Rev.\  D {\bf 23}, 1880 (1981).

\bibitem{Chivukula:1992pm}
  R.~S.~Chivukula, M.~Golden and E.~H.~Simmons,
  Phys.\ Rev.\ Lett.\  {\bf 70}, 1587 (1993)
  [arXiv:hep-ph/9210276].

\bibitem{Appelquist:1995en}
  T.~Appelquist, M.~Schwetz and S.~B.~Selipsky,
  Phys.\ Rev.\  D {\bf 52}, 4741 (1995)
  [arXiv:hep-ph/9502387].

\bibitem{Khlebnikov:1995qb}
  S.~Y.~Khlebnikov and R.~G.~Schnathorst,
  Phys.\ Lett.\  B {\bf 358}, 81 (1995)
  [arXiv:hep-ph/9504389].





\bibitem{Weinberg:1979bn}
  S.~Weinberg,
  Phys.\ Rev.\  D {\bf 19}, 1277 (1979).
%

\bibitem{Susskind:1978ms}
  L.~Susskind,
  Phys.\ Rev.\  D {\bf 20}, 2619 (1979).

\bibitem{Eichten:1979ah}
  E.~Eichten and K.~D.~Lane,
  Phys.\ Lett.\  B {\bf 90}, 125 (1980).


\bibitem{Holdom:1981rm}
  B.~Holdom,
  Phys.\ Rev.\  D {\bf 24}, 1441 (1981).
  
\bibitem{Holdom:1984sk}
  B.~Holdom,
  Phys.\ Lett.\  B {\bf 150}, 301 (1985).

\bibitem{Yamawaki:1985zg}
  K.~Yamawaki, M.~Bando and K.~i.~Matumoto,
  Phys.\ Rev.\ Lett.\  {\bf 56}, 1335 (1986).
  
\bibitem{Appelquist:1986an}
  T.~W.~Appelquist, D.~Karabali and L.~C.~R.~Wijewardhana,
  Phys.\ Rev.\ Lett.\  {\bf 57}, 957 (1986).
  
\bibitem{Appelquist:1986tr}
  T.~Appelquist and L.~C.~R.~Wijewardhana,
  Phys.\ Rev.\  D {\bf 35}, 774 (1987).

\bibitem{Appelquist:1987fc}
  T.~Appelquist and L.~C.~R.~Wijewardhana,
  Phys.\ Rev.\  D {\bf 36}, 568 (1987).

\bibitem{Sannino:2004qp}
  F.~Sannino and K.~Tuominen,
  Phys.\ Rev.\  D {\bf 71}, 051901 (2005)
  [arXiv:hep-ph/0405209].

\bibitem{Dietrich:2006cm}
  D.~D.~Dietrich and F.~Sannino,
  Phys.\ Rev.\  D {\bf 75}, 085018 (2007)
  [arXiv:hep-ph/0611341].

%
 


\bibitem{Dolan:1973qd}
  L.~Dolan and R.~Jackiw,
  Phys.\ Rev.\  D {\bf 9}, 3320 (1974).
  
\bibitem{Weinberg:1974hy}
  S.~Weinberg,
  Phys.\ Rev.\  D {\bf 9}, 3357 (1974).

\bibitem{Fendley:1987ef}
  P.~Fendley,
  Phys.\ Lett.\  B {\bf 196}, 175 (1987).
  
\bibitem{Espinosa:1992gq}
  J.~R.~Espinosa, M.~Quiros and F.~Zwirner,
  Phys.\ Lett.\  B {\bf 291}, 115 (1992)
  [arXiv:hep-ph/9206227].

\bibitem{Parwani:1991gq}
  R.~R.~Parwani,
  Phys.\ Rev.\  D {\bf 45}, 4695 (1992)
  [Erratum-ibid.\  D {\bf 48}, 5965 (1993)]
  [arXiv:hep-ph/9204216].

\bibitem{Anderson:1991zb}
  G.~W.~Anderson and L.~J.~Hall,
  Phys.\ Rev.\  D {\bf 45}, 2685 (1992).
  
\bibitem{Carrington:1991hz}
  M.~E.~Carrington,
  Phys.\ Rev.\  D {\bf 45}, 2933 (1992).

\bibitem{Dine:1992wr}
  M.~Dine, R.~G.~Leigh, P.~Y.~Huet, A.~D.~Linde and D.~A.~Linde,
  Phys.\ Rev.\  D {\bf 46}, 550 (1992)
  [arXiv:hep-ph/9203203].
  
\bibitem{Arnold:1992rz}
  P.~Arnold and O.~Espinosa,
  Phys.\ Rev.\  D {\bf 47}, 3546 (1993)
  [Erratum-ibid.\  D {\bf 50}, 6662 (1994)]
  [arXiv:hep-ph/9212235].

\bibitem{Fodor:1994bs}
  Z.~Fodor and A.~Hebecker,
  Nucl.\ Phys.\  B {\bf 432}, 127 (1994)
  [arXiv:hep-ph/9403219].
  

\bibitem{Manton:1983nd}
  N.~S.~Manton,
  Phys.\ Rev.\  D {\bf 28}, 2019 (1983).
  
\bibitem{Klinkhamer:1984di}
  F.~R.~Klinkhamer and N.~S.~Manton,
  Phys.\ Rev.\  D {\bf 30}, 2212 (1984).
  

%


\bibitem{Iwasaki:2003de}
  Y.~Iwasaki, K.~Kanaya, S.~Kaya, S.~Sakai and T.~Yoshie,
  Phys.\ Rev.\  D {\bf 69}, 014507 (2004)
  [arXiv:hep-lat/0309159].

\bibitem{AliKhan:2000iz}
  A.~Ali Khan {\it et al.}  [CP-PACS Collaboration],
  Phys.\ Rev.\  D {\bf 63}, 034502 (2001)
  [arXiv:hep-lat/0008011].
    
\bibitem{Damgaard:1997ut}
  P.~H.~Damgaard, U.~M.~Heller, A.~Krasnitz and P.~Olesen,
  Phys.\ Lett.\  B {\bf 400}, 169 (1997)
  [arXiv:hep-lat/9701008].

\bibitem{Maezawa:2007fd}
  Y.~Maezawa, S.~Aoki, S.~Ejiri, T.~Hatsuda, N.~Ishii, K.~Kanaya and N.~Ukita,
  J.\ Phys.\ G {\bf 34}, S651 (2007)
  [arXiv:hep-lat/0702005].

\bibitem{Ukita:2006pc}
  N.~Ukita, S.~Ejiri, T.~Hatsuda, N.~Ishii, Y.~Maezawa, S.~Aoki and K.~Kanaya,
  PoS {\bf LAT2006}, 150 (2006)
  [arXiv:hep-lat/0610038].


\bibitem{Appelquist:2003hn}
  T.~Appelquist, M.~Piai and R.~Shrock,
  Phys.\ Rev.\  D {\bf 69}, 015002 (2004)
  [arXiv:hep-ph/0308061].



\bibitem{Yao:2006px}
  W.~M.~Yao {\it et al.}  [Particle Data Group],
  J.\ Phys.\ G {\bf 33}, 1 (2006).

\bibitem{Dugan:1991ck}
  M.~J.~Dugan and L.~Randall,
  Phys.\ Lett.\  B {\bf 264}, 154 (1991).

\bibitem{Peskin:2001rw}
  M.~E.~Peskin and J.~D.~Wells,
  Phys.\ Rev.\  D {\bf 64}, 093003 (2001)
  [arXiv:hep-ph/0101342].



\bibitem{Banks:1981nn}
  T.~Banks and A.~Zaks,
  Nucl.\ Phys.\  B {\bf 196}, 189 (1982).
  

\bibitem{Appelquist:1996dq}
  T.~Appelquist, J.~Terning and L.~C.~R.~Wijewardhana,
  Phys.\ Rev.\ Lett.\  {\bf 77}, 1214 (1996)
  [arXiv:hep-ph/9602385].

\bibitem{Miransky:1996pd}
  V.~A.~Miransky and K.~Yamawaki,
  Phys.\ Rev.\  D {\bf 55}, 5051 (1997)
  [Erratum-ibid.\  D {\bf 56}, 3768 (1997)]
  [arXiv:hep-th/9611142].
  
\bibitem{Chivukula:1996kg}
  R.~S.~Chivukula,
  Phys.\ Rev.\  D {\bf 55}, 5238 (1997)
  [arXiv:hep-ph/9612267].
  
\bibitem{Appelquist:1998rb}
  T.~Appelquist, A.~Ratnaweera, J.~Terning and L.~C.~R.~Wijewardhana,
  Phys.\ Rev.\  D {\bf 58}, 105017 (1998)
  [arXiv:hep-ph/9806472].


\bibitem{Harada:2003dc}
  M.~Harada, M.~Kurachi and K.~Yamawaki,
  Phys.\ Rev.\  D {\bf 68}, 076001 (2003)
  [arXiv:hep-ph/0305018].

  
\bibitem{Kurachi:2006mu}
  M.~Kurachi and R.~Shrock,
  Phys.\ Rev.\  D {\bf 74}, 056003 (2006)
  [arXiv:hep-ph/0607231].

\bibitem{Kurachi:2006ej}
  M.~Kurachi and R.~Shrock,
  JHEP {\bf 0612}, 034 (2006)
  [arXiv:hep-ph/0605290].
  
\bibitem{Bando:1984ej}
  M.~Bando, T.~Kugo, S.~Uehara, K.~Yamawaki and T.~Yanagida,
  Phys.\ Rev.\ Lett.\  {\bf 54}, 1215 (1985).
  
\bibitem{Bando:1987br}
  M.~Bando, T.~Kugo and K.~Yamawaki,
  Phys.\ Rept.\  {\bf 164}, 217 (1988).

\bibitem{Harada:1993jk}
  M.~Harada, T.~Kugo and K.~Yamawaki,
  Phys.\ Rev.\ Lett.\  {\bf 71} (1993) 1299
  [arXiv:hep-ph/9303257].

\bibitem{Harada:1993qi}
  M.~Harada, T.~Kugo and K.~Yamawaki,
  Prog.\ Theor.\ Phys.\  {\bf 91}, 801 (1994)
  [arXiv:hep-ph/9303258].
  
\bibitem{Harada:2003jx}
  M.~Harada and K.~Yamawaki,
  Phys.\ Rept.\  {\bf 381}, 1 (2003)
  [arXiv:hep-ph/0302103].
  
\bibitem{Harada:2000kb}
  M.~Harada and K.~Yamawaki,
  Phys.\ Rev.\ Lett.\  {\bf 86}, 757 (2001)
  [arXiv:hep-ph/0010207].
  
\bibitem{Harada:2001it}
  M.~Harada and C.~Sasaki,
  Phys.\ Lett.\  B {\bf 537}, 280 (2002)
  [arXiv:hep-ph/0109034].
   
  
\bibitem{Hasenbusch:2001ne}
  M.~Hasenbusch,
  Phys.\ Lett.\  B {\bf 519}, 177 (2001)
  [arXiv:hep-lat/0107019].

\bibitem{Luscher:2003vf}
  M.~Luscher,
  JHEP {\bf 0305}, 052 (2003)
  [arXiv:hep-lat/0304007].
  
\bibitem{Luscher:2005rx}
  M.~Luscher,
  Comput.\ Phys.\ Commun.\  {\bf 165}, 199 (2005)
  [arXiv:hep-lat/0409106].

\bibitem{Kuramashi:2006np}
  Y.~Kuramashi {\it et al.}  [PACS-CS Collaboration],
  PoS {\bf LAT2006}, 029 (2006)
  [arXiv:hep-lat/0610063].


\bibitem{van den Eshof:2002ms}
  J.~van den Eshof, A.~Frommer, T.~Lippert, K.~Schilling and H.~A.~van der Vorst,
  Comput.\ Phys.\ Commun.\  {\bf 146}, 203 (2002)
  [arXiv:hep-lat/0202025].

\bibitem{Frommer:1995ik}
  A.~Frommer, B.~Nockel, S.~Gusken, T.~Lippert and K.~Schilling,
  Int.\ J.\ Mod.\ Phys.\  C {\bf 6}, 627 (1995)
  [arXiv:hep-lat/9504020].

\bibitem{Chiu:2002eh}
  T.~W.~Chiu, T.~H.~Hsieh, C.~H.~Huang and T.~R.~Huang,
  Phys.\ Rev.\  D {\bf 66}, 114502 (2002)
  [arXiv:hep-lat/0206007].
  
\bibitem{Cundy:2004pza}
  N.~Cundy, J.~van den Eshof, A.~Frommer, S.~Krieg, T.~Lippert and K.~Schafer,
  Comput.\ Phys.\ Commun.\  {\bf 165}, 221 (2005)
  [arXiv:hep-lat/0405003].


\bibitem{Vranas:2006zk}
  P.~M.~Vranas,
  Phys.\ Rev.\  D {\bf 74}, 034512 (2006)
  [arXiv:hep-lat/0606014].
  
\bibitem{Fukaya:2006vs}
  H.~Fukaya, S.~Hashimoto, K.~I.~Ishikawa, T.~Kaneko, H.~Matsufuru, T.~Onogi and N.~Yamada
                  [JLQCD Collaboration],
  Phys.\ Rev.\  D {\bf 74}, 094505 (2006)
  [arXiv:hep-lat/0607020].

\cite{Fukaya:2007fb}
\bibitem{Fukaya:2007fb}
  H.~Fukaya {\it et al.}  [JLQCD Collaboration],
  Phys.\ Rev.\ Lett.\  {\bf 98}, 172001 (2007)
  [arXiv:hep-lat/0702003].
  
\bibitem{Basile:2004wa}
  F.~Basile, A.~Pelissetto and E.~Vicari,
  JHEP {\bf 0502}, 044 (2005)
  [arXiv:hep-th/0412026].

\bibitem{Karsch:1998qj}
  F.~Karsch and M.~Lutgemeier,
  Nucl.\ Phys.\  B {\bf 550}, 449 (1999)
  [arXiv:hep-lat/9812023].
  
\bibitem{Engels:2005te}
  J.~Engels, S.~Holtmann and T.~Schulze,
  Nucl.\ Phys.\  B {\bf 724}, 357 (2005)
  [arXiv:hep-lat/0505008].
  
\bibitem{Catterall:2007yx}
  S.~Catterall and F.~Sannino,
  Phys.\ Rev.\  D {\bf 76}, 034504 (2007)
  [arXiv:0705.1664 [hep-lat]].



\end{thebibliography}

\end{document}